\documentstyle[11pt]{article}

\begin{document}
\newcommand{\be}{\begin{equation}}
\newcommand{\ee}{\end{equation}}
\newcommand{\beqa}{\begin{eqnarray}}
\newcommand{\eeqa}{\end{eqnarray}}
 \mathchardef\lag="124C
\def\slash{\rlap{/}}
\newcommand{\EE}{$e^+e^- \:$}
\newcommand{\lH}{$lH$}
\newcommand{\ph}{p_{h}}
\newcommand{\Eh}{E_{h}}
\newcommand{\Ep}{E^{\prime}}
\newcommand{\pht}{{{\mbox{ \boldmath $p$}}}_{h\perp}}
\newcommand{\php}{p_{h\,\parallel}}
\newcommand{\bp}{\mbox{\boldmath $p$}}
\newcommand{\bxt}{\mbox{\boldmath $x$}_{\perp}}
\newcommand{\bk}{\mbox{\boldmath $k$}}
\newcommand{\bpt}{\mbox{\boldmath $p$}_{\perp}}
\newcommand{\bkt}{\mbox{\boldmath $k$}_{\perp}}
\newcommand{\xb}{x_{\scriptscriptstyle B}}
\newcommand{\ddp}[2]{\frac{d^{\, 3} #1}{(2\pi)^3 2 #2}}
\newcommand{\dtp}{d\tilde{p}}
\newcommand{\cw}{{\cal W}}
\newcommand{\ch}{{\cal H}}
\newcommand{\cd}{{\cal D}}
\newcommand{\pha}{\bp_{h'}}
\newcommand{\chb}{\overline{\cal H}}
\newcommand{\hb}{\overline{ H}}
\newcommand{\ci}{{\cal I}}
\newcommand{\phti}[1]{p_{h\perp #1}}
\newcommand{\tot}{\frac{\theta}{2}}
\newcommand{\som}{\sum_h \int d^{\, 3}p_h}
\newcommand{\mhk}{m_{h}^{2}}
\newcommand{\nn}{\nonumber}
\newcommand{\slp}{\slash p   }
\newcommand{\slq}{\slash q  }
\newcommand{\slk}{\slash k }
\newcommand{\slP}{\slash P  }
\newcommand{\slH}{\slash H  }
\newcommand{\pp}{p^{+}}
\newcommand{\pmin}{p^{-}}
\newcommand{\ptr}{{\mbox{ \boldmath $p$}}_{\perp}}
\newcommand{\kt}{{\mbox{ \boldmath $k$}}_{\perp}}
\newcommand{\kp}{k^{+}}
\newcommand{\km}{k^{-}}
\newcommand{\qp}{q^{+}}
\newcommand{\cf}{\overline{\cal F}}
\newcommand{\qm}{q^{-}}
\newcommand{\Pp}{P^{+}}
\newcommand{\Pm}{P^{-}}
\newcommand{\phm}{p_{h}^{-}}
\newcommand{\phplus}{p_{h}^{+}}
\newcommand{\xp}{x^{+}}
\newcommand{\xm}{x^{-}}
\newcommand{\xt}{{\mbox{ \boldmath $x$}}_{\perp}}
\newcommand{\gp}{\gamma^{+}}
\newcommand{\gm}{\gamma^{-}}
\newcommand{\gt}{{\mbox{ \boldmath $\gamma$}}_{\perp}}
\newcommand{\wt}{\sqrt{2}}
\newcommand{\DP}{\frac{1}{2m}\frac{1}{(2\pi)^4} \frac{1}{2 \Eh}
\int \ddp{P_R}{E_R}\,\int \ddp{P_X}{E_X}\,}
\newcommand{\G}[1]{\Gamma^{#1}}
\newcommand{\Gt}[1]{\widetilde{\Gamma}^{#1}}
\newcommand{\psib}{\overline{\psi}}
\newcommand{\xint}{\int d^{\,4}x\, e^{iq\cdot x}}
\newcommand{\xib}{\overline{\xi}}
\newcommand{\ah}{a_h}
\newcommand{\ahk}{a_{h}^{\dagger}}
\newcommand{\gv}{\gamma^{5}}
\newcommand{\gmu}{\gamma^{\mu}}
\newcommand{\gnu}{\gamma^{\nu}}
\newcommand{\gmn}{g^{\mu\nu}}
\newcommand{\Wmn}{W_{\mu\nu}}
\newcommand{\cwmn}{\cw_{\mu\nu}}
\newcommand{\bra}{<\! }
\newcommand{\ket}{\! >}
\newcommand{\wf}{\cw_{1, \perp =0}}
\newcommand{\ub}{\overline{u}}
\newcommand{\vb}{\overline{v}}
\newcommand{\tr}{\mbox{Tr}}
\newcommand{\wb}{\overline{W}}
\newcommand{\sig}{\sigma_{e^+ e^-}}
\newcommand{\cwb}{\overline{{\cal W}}}
\newcommand{\fb}{\overline{F}}
\newcommand{\np}{\hat n_+}
\newcommand{\nm}{\hat n_-}

\title{\sf  Quark correlation functions in deep inelastic
semi-inclusive processes  }
\author{{\sf J. Levelt and P.J. Mulders\thanks{
\sf  also at Department of Physics and Astronomy, Free University, Amsterdam}}
\\
\mbox{ } \\
{\sf National Institute for Nuclear Physics and High Energy Physics} \\
{\sf (NIKHEF-K), P.O. Box 41882, NL-1009 DB Amsterdam, the Netherlands}}
\date{}
 \maketitle

\begin{abstract}
{\sf  We investigate one-particle semi-inclusive processes in
lepton-hadron scattering. In unpolarized scattering order $Q^{-1}$
corrections appear only when transverse momenta are detected.
We consider the twist two and three matrix elements and calculate
the semi-inclusive structure functions in terms of quark correlation
functions. We find that at twist three level not only the standard quark
distribution and fragmentation function contribute, but also two
new transverse "profile functions". We demonstrate explicit gauge
invariance of the hadronic tensor at the twist three level. The
results of our approach are used to calculate expressions for some
cross sections for semi-inclusive processes.
\newline
 }
\end{abstract}
\vspace{10cm}
\noindent
{\sf NIKHEF 92-P9\\
({\bf Revised version})\\
February 93\\
\vspace{2cm}
\noindent
submitted to Phys. Rev. D}
%\newpage
{\sf

\section{\sf Introduction }
 In this paper we will investigate unpolarized one-particle
semi-inclusive lepton-hadron scattering in the deep inelastic
 scattering (DIS) limit. We will present the analysis
in terms of  quark correlation functions. These objects encode the
non-perturbative parts of the process, which we as yet are not
able to calculate other than using models. We will present an
analysis which is complete up to and including ${\cal O}(Q^{-1})$
where $Q^2$ is the negative square of the virtual photon momentum.
In this order we will need to introduce four different projections
of the quark correlation functions, called "profile functions".
The fourier transform of two of these are respectively the quark
distribution function and the quark fragmentation function which
are known from the naive quark parton model at leading order.
The two new functions arise in ${\cal O}(Q^{-1})$ and do not have
simple parton model interpretations. The treatment of (intrinsic)
transverse momentum in this process will turn out to be essential
for these new  profile functions. Explicit gauge invariance will be
maintained throughout the calculation, as we will discuss now in
more detail.

As in all lepton scattering experiments in the one-photon-exchange
approximation the semi-inclusive cross section will be essentially given
by a contraction between a leptonic tensor and a hadronic tensor. All the
interesting physics resides in the hadronic tensor. For the one-particle
semi-inclusive cross section the hadronic tensor is given by (see
section 2 for exact definitions and conventions)
\be
2M\cw_{\mu\nu} =  \frac{1}{(2\pi)^4}  \int \ddp{P_X}{E_X}\, \int
                    d^{\,4}x\,
                     e^{iq \cdot x}
                  \:
                   <\! P|J_{\nu}(x)|P_X , \ph  \!> <\! \ph , P_X
	       	 |J_{\mu}(0)|P\!>.
\ee
Here $q$ is the virtual photon momentum, $P$ and $p_h$ are nucleon
and observed hadron momentum and $P_X$ is shorthand for the ensemble of
unobserved hadrons in the final state. Also the phase space integral over
$P_X$ should be understood to be a summation over all $n$-particle states of
all types. We denote this tensor by the diagram in figure \ref{bigblobs}a.
In terms of quarks and gluons we know the photon current, which only couples
to quarks. This gives us an expression for the hadronic tensor in terms of
quark
fields
\beqa
2M\cw_{\mu\nu} & = & \frac{1}{(2\pi)^4}  \int \ddp{P_X}{E_X}\, \int
                    d^{\,4}x\,
                     e^{iq \cdot x}   \nonumber \\
                & & \times \:
                   <\! P|\psib (x) \gamma_\nu \psi (x)|P_X , \ph  \!>
                   <\! \ph , P_X
	       	   |\psib (0) \gamma_\mu \psi(0)|P\!>,
\eeqa
depicted by the diagram in figure \ref{bigblobs}b. Flavour degrees of freedom
are suppressed throughout. Because of the colour and electromagnetic gauge
invariance of these quark currents   the hadronic tensor is gauge invariant
and $q_\mu \cw^{\mu\nu}=0$. We will make a diagrammatic expansion of which
the first term is the "Born" diagram, i.e. the naive parton model, and
the following diagrams have successively more gluons or quarks participating.
After a specific choice of gauge this diagrammatic expansion will   be the same
as an expansion in $Q^{-1}$. The relationship between diagrams with
successively more gluons  or quarks participating and their order in $Q^{-1}$
or "twist" is
discussed at great length by Jaffe and Ji in \cite{JaffeJi}. The separation of
a diagram in a hard scattering part and soft correlation functions is discussed
by Ellis, Furmanski and Petronzio (EFP) \cite{EFP} and by Collins et al. in
\cite{Collinsbook}. Qiu \cite{Qiu} and Ji \cite{Ji} among others discuss this
approach  for inclusive scatterings. Several authors have used these methods
for polarized lepton-hadron and polarized Drell-Yan scattering
\cite{JaffeJi,pol}. In the same spirit
we will use this method for semi-inclusive lepton-hadron scattering. The Born
term, shown in figure \ref{born}, is given by
\beqa
2M\cw^{B\mu\nu} & = & \frac{1}{(2\pi)^4}   \xint
                     \: \gamma^{\nu}_{ij}\gamma^{\mu}_{kl}  \nn \\
                 && \times\left[<\!P|\psib_i (x) \psi_l (0)|P\! >
                    <\!0|\psi_j (x) \ahk \ah \psib_k (0)|0\! >
                      \right. \nn \\
                 && +\left.<\!P|\psi_j (x) \psib_k (0)|P\! >
                    <\!0|\psib_i (x) \ahk \ah\psi_l (0)|0\! >
                     \right]  .
\label{plas}
\eeqa
Several comments are in order here. We now have two separate, unobserved final
states which are integrated over using completeness, represented by the top
and bottom parts of the diagram in figure \ref{born}.  In the top part of the
diagram the additional hadron in the final state appears. In appendix A it is
shown how to handle this when integrating over the final state. The
result is that we are left with a hadron number operator. In the diagram we
see  the separation in   quark correlation functions representing the soft part
and the two gamma matrices that denote the hard photon-quark scattering.
Furthermore we note that eq. (\ref{plas}) contains contributions of quarks and
antiquarks. In the rest of this paper
we will not explicitly account for the antiquarks since they behave exactly
like the quarks in a distinct sector. Of course the numerical values of quark
and antiquark correlation functions can be quite different. Finally we note
that
we will be concerned with hadrons produced from the struck quark, i.e.
belonging
 to the (isolated) current jet. This will pose restrictions on the scaling
variables $\xb$ and $z$, as will become clear later in the discussion of
kinematics and factorization.

Evaluating eq. (\ref{plas})  we will find the following contributions
\be
\cw^{B}_{\mu\nu}(Q) = w^{B0}_{\mu\nu} + w^{B1}_{\mu\nu}\cdot \frac{1}{Q}
			+ {\cal O}\left( \frac{1}{Q^2} \right).
\ee
where $q^\mu w^{B0}_{\mu\nu}=0$ but $q^\mu w^{B1}_{\mu\nu}\neq 0$ so
we explicitly loose gauge invariance in ${\cal O}(Q^{-1})$. This is not
surprising since the quark fields in the matrix elements in eq. (\ref{plas})
are taken at different space-time points, so this matrix element will in
general not be invariant under a local gauge transformation.

We will discuss how one can restore gauge invariance in such a situation before
returning to the diagrammatic expansion. Gauge invariance of a bilocal matrix
element $\psib(x)\psi(x')$ requires a link operator   $\psib (x) L(x,x')
\psi(x')$,
which can be written in the form
\be
     L(x,x')={\cal P}\exp (ig\int\,d^{4}y \ \xi^\alpha (x-y,x'-y) A_\alpha (y)
),
\ee
where $A$ is the appropriate gauge field \cite{Peierls}.  Gauge invariance is
restored when $\xi$ satisfies
\be
	\frac{\partial}{\partial y^\alpha} \xi^\alpha (x-y,x'-y) =
	\delta^4 (x'-y)-\delta^4 (x-y).
	\label{cond}
\ee
The path ordering ${\cal P}$ is then defined along the flow lines of the
"velocity"-field $\xi^\alpha$.
  A special choice of $\xi$ leads to the well known line integral
\be
	L(x,x')={\cal P}\exp (ig\int^{x}_{x'} dy^\alpha A_\alpha (y) ).
\ee
So if one expands in orders of $g$, which is in our case also an
expansion\footnote{\sf This is only true in a specific gauge, we will discuss
this later on.}
in orders of $Q^{-1}$ one concludes that restoring gauge invariance requires
addition of a term
\be
	ig \psib (x) \int d^4 y \ \xi^\alpha (x-y,-y) A_\alpha (y) \psi (0)
	\label{addterm}
\ee
to the Born term.

We return to our diagrammatic expansion, and inspired by eq. (\ref{addterm}),
consider in addition to the Born term the four diagrams of figure \ref{gauge}.
They have one fermion propagator in the hard scattering part and their
leading contribution is ${\cal O}(Q^{-1})$.
Generically
\be
	\cw_{\mu\nu}^{Bg}(Q) = w_{\mu\nu}^{Bg1}\cdot \frac{1}{Q} + {\cal O}
	\left( \frac{1}{Q^2} \right).
\ee
The essential part of our proof of gauge invariance will be as follows.
We will show that in these diagrams the soft parts are matrix elements of
operators of the form in  eq. (\ref{addterm}) and in this way identify
$\xi^\alpha$
in our case. Then we will show that this $\xi^\alpha$ respects the condition
eq. (\ref{cond})   up to ${\cal O} (Q^{-2})$.
We claim that all other diagrams with extra gluons either give rise
to the logs of the radiative corrections, or are suppressed by powers
of $Q$, or ultimately vanish in a suitably chosen gauge.

Summarizing we find
\be
	\cw_{\mu\nu} = w^{B0}_{\mu\nu}+(w^{B1}_{\mu\nu}+w^{Bg1}_{\mu\nu})
	\cdot \frac{1}{Q} + {\cal O} \left(\frac{1}{Q^2} \right)
\ee
and $q^\mu \cw_{\mu\nu}= {\cal O}(Q^{-2})$ in our calculation.   After
explicit evaluation of these five diagrams we express $\cw_{\mu\nu}$ in four
unknown
 profile functions, fourier transforms of specific projections of the
quark correlation functions.

The rest of the paper is structured as follows. In section 2 we discuss
the generalities of the semi-inclusive cross section, kinematics, structure
functions and inclusive reductions. In section 3 we will first look at the
kinematics of the relevant diagrams and discuss what kind of assumptions are
made
related to the factorization into hard scattering and correlation functions and
make a twist-analysis,
then we will define the basics objects of which our diagrams are  composed and
show how they relate to the four profile functions which characterize the
process. Then in section 4 we will present the actual calculation of the
hadronic
tensor and demonstrate  gauge invariance. Finally in section 5 we present our
results as expressions for various cross section using these four scalar
functions.

\section{\sf  Semi-inclusive deep inelastic lepton hadron scattering}
We will consider semi-inclusive processes of the type $eH\rightarrow
e'hX$, where $H$ is a hadronic target with mass $M$, $h$ is a hadron
with mass $m_h$ detected in coincidence with the scattered electron and $X$ is
 the rest of the final state. The momenta are defined in figure \ref{bigblobs},
the angles in the target rest frame (TRF) are defined in figure \ref{angles}.
In a one-hadron semi-inclusive process we can form four independent invariants
 $q^2$, $P\cdot q$, $P\cdot p_h$ and $p_h\cdot q$. This includes the invariants
that one is familiar with in inclusive scattering:
\beqa
Q^2 &=& -q^2 \stackrel{\mbox{\scriptsize TRF}}{=}
4EE' \sin^2 \left( \frac{\theta}{2} \right),\\
\nu &=& \frac{P\cdot q}{M} \stackrel{\mbox{\scriptsize TRF}}{=}
E-E',
\eeqa
and the ratios
\be
\xb = \frac{Q^2}{2M\nu}, \mbox{\hspace{2cm}}
y=\frac{P\cdot q}{P\cdot k}=\frac{\nu}{E}.
\ee
For semi-inclusive scattering one or more particles (=momenta) in the final
state are measured.   The  additional  invariants are fixed by the energy $E_h$
and the component of $p_h$ along {\boldmath$ q$},indicated by
$\php$, or equivalently by $E_h$ and $\pht^2$. In the TRF they do not
depend on the azimuthal angle $\phi_h$. Note that if more than one momentum
in the final state is measured the relative azimuthal angle appears in the
invariant $p_1 \cdot p_2$. In particular in the TRF one has
${\mbox{\boldmath $p$}}_{1\perp} \cdot{\mbox{\boldmath $p$}}_{2\perp}=
|{\mbox{\boldmath $p$}}_{1\perp} | |{\mbox{\boldmath $p$}}_{2\perp}|
\cos (\phi_1 - \phi_2)$. In DIS processes the ratio
\be
z=\frac{P\cdot p_h}{P\cdot q} \stackrel{\mbox{\scriptsize TRF}}{=}
\frac{E_h}{\nu}
\ee
will be useful.
We will consider in this paper the limit of deep inelastic scattering, in which
$Q^2 \rightarrow \infty$ and $\nu \rightarrow \infty$, keeping the ratio $x_B$
fixed.
In this limit\footnote{\sf $q^\pm=
\frac{1}{\sqrt{2}}(q^0 \pm q^3)$.} (defining $\hat{z}$ along
$-{\mbox{\boldmath $q$}}$), $\qm \rightarrow \infty$. One then immediately sees
that
 for a particle produced in the current jet the other invariant involving $p_h$
and $q$ with $p_h$ on mass shell is:
\be
\frac{2p_h \cdot q}{q^2} \simeq \frac{p_{h}^{-}\qp}{\qm\qp} =
\frac{p_{h}^{-}}{\qm} \simeq z.
\ee
The cross section for the semi-inclusive process is given by
\be
\frac{2E_h d\sigma}{d^3 p_h d\Omega dE'} =\frac{\alpha^2}{Q^4} \frac{E'}{E}
L_{\mu\nu}\cw^{\mu\nu},
\ee
or
 \be
\frac{2E_h d\sigma}{d^3 p_h d\xb dy} = \frac{\pi \alpha^2
2My}{Q^4} L_{\mu\nu}\cw^{\mu\nu}.
\ee
where $L_{\mu\nu}$ is the spin averaged lepton tensor
\be
L_{\mu\nu}=2 k_\mu k_\nu^\prime  + 2k_\nu k_\mu^{\prime} -Q^2 g_{\mu\nu}
\ee
and the hadronic
tensor is the product of non-elementary hadronic currents,
\beqa
2 M\cw_{\mu\nu} & = & \frac{1}{(2\pi)^4}   \int \ddp{P_X}{E_X}
<\! P|J_{\nu}(0)|P_X , \ph  \!> <\! \ph , P_X |J_{\mu}(0)|P\!>\!  \nonumber \\
 & & \mbox{\hspace{3cm}}\times \: (2 \pi)^4 \delta^{4}
\,(P+q-\ph -P_X).  \label{hadronictensor1}
\eeqa
The integral over $P_X$ indicates a complete summation over all multiparticle
states
of all hadron types. We have split off the invariant phase space for the
detected
hadron, which can also be written:
\be
\frac{d^3 p_h}{2 E_h} = \frac{dE_h  d \! \pht^2 d\phi_h}{4\php} =
\frac{1}{4} \frac{\nu}{\php} dz  d \! \pht^2 d\phi_h.
\ee
 The most
general gauge invariant, Lorentz covariant hadronic tensor  for
the electromagnetic process
(parity and time reversal invariant) can be decomposed in four structure
functions
$\cw_i (x,Q^2,z,\pht^2)$  \cite{piet}:
\beqa
\cw_{\mu\nu} (P, \ph , q ) & = & (\frac{q_\mu q_\nu}{q^2}-g_{\mu
\nu})\:\cw_1 + \frac{T_\mu T_\nu}{M^2}\:\cw_2  \nonumber \\
 & & +\frac{\phti{\mu}
T_{\nu}+T_{\mu} \phti{\nu}}{M m_h} \:\cw_3 +
\frac{\phti{\mu}\phti{\nu}}{m_{h}^{2}}\:\cw_4 ,
\label{generaltensor}
\eeqa
where
\be
T_{\mu}=P_{\mu} - \frac{(P\cdot q)}{q^2} q_\mu,
\ee
and
\be
p_{h\perp}^\mu \stackrel{\mbox{\scriptsize TRF}}{=} (0, \pht, 0).
\ee
As in the inclusive case, it is convenient to introduce photon polarizations.
The
structure functions can be expressed in terms of:
\beqa
&&\cw_L  =  \epsilon_{L} \cdot \cw \cdot \epsilon_{L}
= -\cw_1 + \frac{{\mbox{\boldmath $q$}}^2}{Q^2}\,\cw_2 , \nonumber \\
&&\cw_{T}  =  \frac{1}{2}\left(\epsilon_{x} \cdot \cw \cdot \epsilon_{x} +
\epsilon_{y} \cdot \cw \cdot \epsilon_{y}\right)
= \cw_1 + \frac{\pht^2}{2m_{h}^{2}}\,\cw_4 , \nonumber \\
&&\cw_{LT} \cos (\phi_h  ) = -( \epsilon_{x} \cdot \cw \cdot \epsilon_{L} +
\epsilon_{L} \cdot \cw \cdot \epsilon_{x})
= -2 \frac{|{\mbox{\boldmath $q$}}|}{Q}
\frac{|\pht |}{m_h} \cos (\phi_h )\,\cw_3 , \nonumber \\
&&\cw_{TT} \cos (2\phi_h)  =  \frac{1}{2}\left(\epsilon_{x} \cdot \cw \cdot
\epsilon_{x}
-\epsilon_{y} \cdot \cw \cdot \epsilon_{y}\right)
=\frac{\pht^2}{2m_{h}^{2}} \cos (2\phi_h )\, \cw_4 .
\label{WLT}
\eeqa
The $\epsilon$'s here are the standard   transversal and longitudinal photon
polarization vectors.
The tensors multiplying these structure functions in
\be
\cw_{\mu\nu} (P, \ph , q )=
P^L_{\mu\nu}\cw_L + P^T_{\mu\nu}\cw_T - \frac{1}{2} P^{LT}_{\mu\nu}
\cw_{LT} + P^{TT}_{\mu\nu} \cw_{TT}
\ee
are
\beqa
P^L_{\mu\nu}&=& \frac{T_\mu T_\nu}{T^2}, \nn \\
P^T_{\mu\nu}&=& -\gmn + \frac{q_\mu q_\nu}{q^2} + \frac{T_\mu T_\nu}{T^2},\nn
\\
P^{LT}_{\mu\nu}&=& \frac{\phti{\mu}
T_{\nu}+T_{\mu} \phti{\nu}}{T |\pht|}, \nn \\
P^{TT}_{\mu\nu}&=& \gmn - \frac{q_\mu q_\nu}{q^2} - \frac{T_\mu T_\nu}{T^2} +2
\frac{\phti{\mu}\phti{\nu}}{\pht^{2}}. \label{projectors}
\eeqa
They can be used to project out the structure functions $\cw_i$
\beqa
\cw_L &=& P^L_{\mu\nu}\cw^{\mu\nu}, \nn \\
\cw_T &=& \frac{1}{2}P^T_{\mu\nu}\cw^{\mu\nu}, \nn \\
\cw_{LT}&=& P^{LT}_{\mu\nu}\cw^{\mu\nu}, \nn \\
\cw_{TT}&=& \frac{1}{2} P^{TT}_{\mu\nu}\cw^{\mu\nu}
\eeqa
as they are orthogonal to each other.

In DIS it is convenient to use the combinations $\ch_i=\ch_i(x,Q^2,z,\pht^2)$,
\beqa
2z\ch_1 &=& M(\cw_1 + \frac{\pht^2}{2m_h^2}\cw_4)=M\cw_T, \nn \\
2z\ch_2 &=& \nu \left( \cw_2 + \frac{\pht^2 Q^2}{2m_h^2 {\mbox{\boldmath
$q$}}^2} \cw_4
\right) =
\frac{\nu Q^2}{{\mbox{\boldmath $q$}}^2} ( \cw_L + \cw_T), \nn \\
2z\ch_3 &=& -\frac{Q^2}{m_h} \cw_3 = \frac{Q^3}{2 |{\mbox{\boldmath $q$}}|
|\pht|}
\cw_{LT}, \nn \\
2z\ch_4 &=& M\xb \frac{Q^2}{m_h^2} \cw_4 =  \frac{Q^4}{{\mbox{\boldmath
$q$}}^2\pht^2}
\cw_{TT}.
\eeqa
For later purposes it is also useful to introduce
\beqa
2z\ch_L &=& M\cw_L \nn \\
&=& 2z\left[ -\ch_1 + \left( 1+ \frac{Q^2}{\nu^2} \right) \frac{\ch_2}
{2\xb}\right] \stackrel{\mbox{\scriptsize DIS}}{=}
2z\left[ -\ch_1 +   \frac{\ch_2}{2\xb}\right].
\eeqa
The cross sections are easily obtained from the contraction of leptonic and
hadronic tensors. Without any approximations one obtains
\beqa
\frac{2E_h d\sigma}{d^3 p_h d\Omega dE'}
  & = &  \left(\frac{d\sigma}{d\Omega}\right)_M \frac{Q^2}{{\mbox{\boldmath
$q$}}^2}
\frac{1}{\epsilon}
\Biggl[ \cw_T +\epsilon \,\cw_L  \nonumber \\
 & &   + \epsilon \,\cw_{TT} \cos(2 \phi_h) + \sqrt{\frac{\epsilon (\epsilon
+1)}{2}}\,\cw_{LT}\cos (\phi_h ) \Biggr] ,
\eeqa
where
\be
\left(\frac{d\sigma}{d\Omega}\right)_M =\frac{4\alpha^2 E^{\prime 2}}{Q^4}
\cos^2 \left( \frac{\theta}{2} \right)
\ee
and
\be
\epsilon^{-1}
= 1+ 2 \frac{{\mbox{\boldmath $q$}}^2}{Q^2} \tan^2 \left( \frac{\theta}{2}
\right).
\ee
It is also possible to observe a second particle in the final state. One then
finds by a similar analysis two extra structure functions associated with the
extra particle, $\cw^2_{LT}$ and $\cw^2_{TT}$, which enter the cross section in
the same way as the one particle structure functions, (see \cite{Donnely}).

Up till here all expressions are completely general. Now we want to specialize
to DIS events. Our attention will be the current jet, considering  particles
produced with bounded transverse momenta, i.e.
\be
m_{h}^2 + \pht^2 \ll \php^2  \simeq    E_h^2.
\label{approx}
\ee
The cross section may be written as (still exact):
\beqa
\frac{d\sigma}{d\xb dz dy  d \! \pht^2 d\phi_h}&=&
\frac{ 4 \pi \alpha^2 M E}{Q^4} \frac{E_h}{\php} \Biggl[
\xb y^2 \ch_1+ \left( 1-y-\frac{M\xb y}{2E} \right) \ch_2  \nn \\
&&  +\frac{|\pht|}{Q}
(2-y) \sqrt{\frac{1-y-\frac{M\xb y}{2E}}{1+\frac{2M\xb}{yE}}}
\cos (\phi_h) \ch_3\nn \\ &&  + \frac{\pht^2}{Q^2} \frac{\left( 1-y-\frac{M\xb
y}{2E}\right)}{1+\frac{2M\xb}{yE}} \cos (2\phi_h ) \ch_4 \Biggr]
\eeqa
which  becomes in the DIS limit
\beqa
 \frac{d\sigma}{d\xb dz dy  d \! \pht^2 d\phi_h}&\stackrel{\mbox{\scriptsize
DIS}}{=}& \frac{ 4 \pi \alpha^2 M E}{Q^4}
\Biggl[ \xb y^2 \,\ch_1 + (1-y) \,\ch_2   \nonumber \\
 & & +\frac{|\pht|}{Q} (2-y) \sqrt{(1-y)}\cos (\phi_h ) \,\ch_3
\nonumber \\
 & &  +\frac{\pht^2}{Q^2} (1-y) \cos( 2 \phi_h ) \,\ch_4 \Biggr].
\label{bergerlang}
\eeqa
Integrating over the azimuthal angle only leaves the first two terms. Often
one is only interested in the $z$-behaviour of the structure functions, so
defining
\beqa
H_i (\xb,Q^2,z) &= &{1\over 2}\int \!  d \! \pht^2 d\phi_h\, \ch_i (\xb, Q^2,z,
\pht^2) \nn \\
&=& \pi \int \!  d \! \pht^2 \ch_i (\xb, Q^2,z, \pht^2)
\eeqa
we are left with
\be
\frac{d\sigma}{d\xb dy dz}\stackrel{\mbox{\scriptsize DIS}}{=}
 \frac{ 8 \pi \alpha^2 M E}{Q^4}\left[ \xb y^2 H_1 + (1-y)
H_2 \right] . \label{xyz}
\ee

In the analysis of experimental data, factorization of the structure functions
is taken to mean that one can write the structure function $H_2$
as a flavour sum over products of quark distribution functions and quark
fragmentation
 functions \cite{Berger}:
\be
H_2(\xb,Q^2,z)=H_2(\xb,z)=\sum_{q} e_{q}^2\, \xb\left[  f_{q/H}(\xb) D_{h/q}
(z)
+f_{\overline{q}/H}(\xb) D_{h/\overline{q}}(z)\right]
, \label{aim}
\ee
where $q$ indicates quarks of different flavours. In this paper we show the
validity
of this factorization in leading order in $Q$, and its extension to ${\cal O}
(Q^{-1})$ and secondly we derive an expression for the fragmentation function
in terms
 of quark   correlation functions. Furthermore one obtains
\be
2\xb \ch_1 (\xb, z, \pht^2) = \ch_2 (\xb, z,\pht^2), \label{CG}
\ee
as $\ch_L \rightarrow 0$ in the DIS limit.

Finally some remarks on the connection to inclusive scattering must be made.
The hadronic tensor for inclusive scattering $W_{\mu\nu} (P,q)$
is given by (see appendix A):
\be
\langle n_h (P,q) \rangle W_{\mu\nu} (P,q)= \int \frac{d^{\,3}p_h}{2E_h}
\cw_{\mu\nu} (P,q,p_h),
\ee
where $\langle n_h (P,q) \rangle$ is the average number of particles produced
of
type $h$ for given $P$ and $q$. Combining the definitions we then have
\beqa
\langle n_h(\xb ) \rangle F_1(\xb ) & = &  \int dz H_1(\xb ,z)   \nn \\
\langle n_h(\xb ) \rangle F_2(\xb ) & = &  \int dz H_2(\xb ,z) ,
\label{efs}
\eeqa
where $F_i$ are the standard dimensionless scaling  structure functions in
inclusive
\lH.

\section{\sf Factorization and correlation functions}
\subsection{\sf  Factorization and twist analysis}
In this section we will discuss the diagrammatic expansion.
We start by considering the kinematics. It is convenient to use lightcone
coordinates for this. With the four vector notation
$p=[p^-, \pp, \bpt]$ with $p^\pm = (p^0 \pm p^3 ) / \sqrt{2}$ we denote
the momenta of the target hadron ($P$), virtual photon ($q$) and final state
 hadron ($p_h$) as
\beqa
	P&=&\left[\frac{\xb M^2 }{A\sqrt{2}}, \frac{A}{\xb\sqrt{2}},
	   {\bf 0}_{\perp} \right] \nn\\
	q&=&\left[\frac{Q^2}{ A\sqrt{2}}, -\frac{ A}{\sqrt{2}},
	{\bf 0}_{\perp}\right]\nn\\
	\ph &=&\left[  \frac{zQ^2}{A\sqrt{2}  },
	\frac{A(\pht^2 +m_{h}^{2}) 	}
	{ z Q^2 \sqrt{2}}, \pht \right] \label{lckinematics}.
\eeqa
Here $A$ is a free parameter fixing the frame among those that are connected by
boosts along the photon-quark direction. For instance $A=\xb M$ corresponds to
the
target rest frame (TRF), while for $\pht=0$, the choice $A=(zQ^2)/m_h$
corresponds
to the  outgoing hadron rest frame. A convenient infinite momentum frame (IMF)
is
found by choosing $A=Q$. Using the momenta $P$ and $q$,
 and the momentum $T$ $\equiv$ $P - (P\cdot q/q^2)\,q$, one can construct the
momenta
\begin{eqnarray}
	\hat T^\mu \equiv \frac{T^\mu}{T} & = & \left[ \frac{Q}{A\sqrt{2}},
	\frac{A}{Q\sqrt{2}}, {\bf 0}_\perp \right], \\
	\hat q^\mu \equiv \frac{q^\mu}{Q} & = & \left[ \frac{Q}{A\sqrt{2}},
	-\frac{A}{Q\sqrt{2}}, {\bf 0}_\perp \right],
\end{eqnarray}
 and the two lightcone unit vectors
\beqa
	\np^\mu &=& \left( \frac{\qm}{Q} \right) (\hat T^\mu - \hat q^\mu), \\
	\nm^\mu &=& \left( -\frac{\qp}{Q} \right) (\hat T^\mu + \hat q^\mu),
\eeqa
with $\np \cdot \nm =1$. Note that in our results the invariants $\xb$ and $z$
are
 used as the ratios
of lightcone coordinates.
Thus the replacements
\beqa
	\xb &\longrightarrow &-\frac{q^+}{P^+}=
	\frac{2\xb}{1+\sqrt{1+\frac{4M^2\xb^2}{Q^2}}}, \nn \\
	z&\longrightarrow &
 	\frac{p_h^-}{q^-}=\frac{z\left(1+\frac{p_{h\parallel}}{E_h}
	\right)}
	{1+\sqrt{1+\frac{4M^2\xb^2}{Q^2}}},
\eeqa
would provide the necessary socalled hadron mass corrections. Furthermore we
assume that for the current jet events we consider, no transverse momenta
growing like $Q$ appear. We consistently neglect terms of order transverse
momentum divided by $Q^2$. Also $\xb$ and $z$ should not be too close to zero.
To be precise, $\xb Q$ and $zQ$ must be at least larger than some hadronic
scale.

Diagrams contributing to the process will be separated into soft hadronic
matrix elements and a hard scattering part. The forward matrix elements
involving the external particles (momenta $P$, $p_h$) are given by
 untruncated (except for external lines) Green functions. Given  a
renormalization scale for these Green functions, they are calculated
as 2-particle irreducible
 in the quark anti-quark channel.  The evolution
between different renormalization scales is then provided by the logarithmic
 corrections discussed below. It is assumed that for all of the Green
functions  the parton
virtualities are restricted to some hadronic scale, i.e. they vanish with
 some inverse power of the virtualities. The relevant diagrams that are
considered
in this paper have been given in figures \ref{born} and \ref{gauge}.
There are other diagrams with one gluon such as those of the type in
figure \ref{gauge}  with the photon-quark-quark vertex and the
gluon-quark-quark
vertex interchanged. Those can be absorbed in the soft parts and will not
 be considered explicitly in our analysis. Diagrams with one gluon which
are important are those shown in figure  \ref{logs}.
They contain in the hard part partons in the final state and will produce
logarithmic corrections from the collinear divergencies\footnote{\sf The
quark self-energy and the photon-quark-quark  vertex correction provide
singular
 terms needed for the mass   factorization.}.
This leads to the evolution of the profile functions $f^+, D^-$, also known
as the Gribov-Lipatov-Altarelli-Parisi equations, as has been proven explicitly
by Ellis, Georgi, Machacek, Politzer and Ross \cite{Ellis}. The non-logarithmic
terms in their calculation are the gluon jet events which appear in order
$\alpha_s$
 and which have been calculated in a naive parton model by K\"{o}nig and Kroll
in ref. \cite{Kroll}. We expect
that from diagrams like those in figure \ref{gauge} which have in addition
gluons in the final state one will obtain $Q^{-1}\log Q^2$ contributions. In
 this paper however, we will not discuss the logarithmic contributions any
further. We do note that in the leading order we can calculate the logarithmic
 corrections in our approach and that the evolution equations are not altered
 by the observation
of transverse momenta. Finally there are diagrams which have a gluon across
between the hadronic parts. With the above assumption on the parton
virtualities
 one immediately sees that they will vanish as some power of $Q^{-2}$.
The general behaviour of the contributions of diagrams with  any number of
partons
 emerging from the hadronic matrix elements is  discussed below.

It is illustrative to first consider the kinematics of the diagrams in figures
\ref{born} and \ref{gauge}. We assign the four momenta
\beqa
	p&=&\left[\frac{\xi (p^2 + \bpt^2)}
	{ A\sqrt{2}}, \frac{ A}{\xi \sqrt{2}} , \bp_{\perp}
  	\right]
	\nn\\
	k&=&\left[\frac{\zeta Q^2}{ A\sqrt{2}},
	\frac{ A  (k^2+\bkt^2) }{
	 \zeta Q^2 \sqrt{2}},{\bk}_{\perp} \right], \nn\\
	p_1&=&\left[ \frac{ (p_1^2+\bp_{1\perp}^2)}{(1-\xi_1)A\sqrt{2}},
	\frac{(1-\xi_1)A}{ \sqrt{2}}, \bp_{1\perp}\right], \nn\\
	k_1&=&\left[\frac{(1-\zeta_1)Q^2}{A\sqrt{2}},\frac{A(k_1^2+
	\bk_{1\perp}^2)}
	{(1-\zeta_1)Q^2\sqrt{2}},\bk_{1\perp}\right].
	\label{kin1}
\eeqa
  Now we apply four momentum conservation $p+q=k$ to derive $\xi=\zeta=1$
in the limit $Q^2\rightarrow \infty$, which is equivalent
to $\pp =-\qp =\xb \Pp$ and $\km =\qm = z\phm$. For later use we also give
expressions for the  momenta in the quark propagators   in figure \ref{gauge},
\beqa
	\frac{p_1+q}{(p_1+q)^2} &=&
	\left[\frac{-1}{\xi_1A\sqrt{2}},
	\frac{A}{Q^2\sqrt{2}}, \frac{-\bp_{1\perp} }
	{\xi_1 Q^2} \right], \nn \\
	\frac{k_1-q}{(k_1-q)^2} &=& \left[\frac{1}{A\sqrt{2}},
 	\frac{-A}{\zeta_1Q^2\sqrt{2}}, \frac{ -\bk_{1\perp}  }
	{\zeta_1 Q^2}  \right].
	\label{kin2}
\eeqa
 It is important to note the suppression of the transverse components relative
to the other components.

 The objects that enter the diagrams in figures \ref{born} and \ref{gauge} are
depicted in figure \ref{rules}. They are partly given by   untruncated (except
for
external particle lines) Green functions for the blobs
 and partly by ordinary QCD Feynman rules.  Suppressing flavour and colour
indices
we have in the hard part the following rules,
\beqa
	\mbox{5a} &:& i\gamma^\nu_{ij}, \\
	\mbox{5b} &:& ig\gamma^\alpha_{ij}, \\
	\mbox{5c} &:& \frac{i}{\slash p},
\eeqa
while the correlation functions are given by
\beqa
	\mbox{5d} &:& T_{ij}(p)=
	\frac{1}{2(2\pi)^4}\int d^{\, 4}x\ e^{-ip\cdot x}
	\bra P |\psib_j(x)\psi_i(0)|P\ket, \\
	\mbox{5e} &:& D_{ji}(k)=\frac{1}{4z(2\pi)^4}
	\int d^{\, 4}x\ e^{ik\cdot x}\bra 0|\psi_j(x)
	\ahk a_h \psib_i(0)|0\ket, \\
	\mbox{5f} &:& F^\alpha_{ij}(p,p_1 ) = \frac{1}{2(2\pi)^8}
	\int d^{\,4}x d^{\, 4}y \ e^{-ip\cdot x} e^{i(p-p_1)\cdot y}
	\bra P|\psib_j(x) A^\alpha(y)\psi_i(0)|P\ket,\\
	\mbox{5g} &:& M^\alpha_{ji}(k,k_1)=\frac{1}{4z(2\pi)^8}
	\int d^{\, 4}xd^{\, 4}y \ e^{ik\cdot x} e^{-i(k-k_1)\cdot y}
	\bra 0|\psi_j (x) A^\alpha (y) \ahk a_h \psib_i(0)|0\ket.
\eeqa
We note that time ordering is not relevant. Using the QCD equations of motion
\cite{Poll} all correlation functions can ultimately be related to functions
containing $\psi_+=(\gamma^-\gamma^+/2)\psi$ and $A^\perp$,
which have c-number (anti)-commutators at equal lightcone time $\xp$
\cite{JaffeTwist4}.

We will now discuss the $Q$-dependence and give the general classification of
all possible  diagrams with more than one parton emerging from the hadronic
matrix element according to the powers of
  $Q$.   We use an extension of the method developed by EFP for inclusive
 scattering. They have proven that for inclusive scattering this method is
equivalent to the Operator Product Expansion (OPE). Because of the hadron in
the
final state we do not have an OPE for semi-inclusive processes. We can,
however,
follow the method of EFP for semi-inclusive processes.

Consider an observable, say one of the structure functions ${\cal H}_i$  in
the expansion of $\cw_{\mu\nu}$. Separating the hard scattering from the soft
pieces we have schematically
\be
	O(P,q,p_h) = \sum_{n,m} O_{n,m}(P,q,p_h),
\ee
where
\be
	O_{n,m}(P,q,p_h) = \int \prod_{i}^{n}\prod_{j}^{m}\,\,dp_i dk_j \
		\hat T (P,p_i) \Gamma (q,p_i,k_i) \hat D (p_h,k_i).
\ee
Omitting the diagrams with partons in the final state, a general term
$O_{n,m}$,
with $n+m$ partons participating in the hard scattering is depicted in figure
\ref{factorization}. To investigate the behaviour of the diagrams for high
$Q^2$
it is convenient to work in the particular infinite momentum frame $A=Q$. Under
the assumption that all parton virtualities must be restricted to some hadronic
scale we can show  as seen explicitly for the momenta appearing in
figures \ref{born} and \ref{gauge}, that $\km_j$ and $\pp_i
\sim Q$ while $\kp_j$ and $p^-_i \sim Q^{-1}$. The transverse momenta
are all of order unity. For the truncated piece $\Gamma$ one can
convince oneself that in the limit $Q \rightarrow \infty$, the $Q$-dependence
 is determined by dimensional arguments. Assume that the canonical dimensions
of the (in general nonlocal) product of operators appearing in $\hat T$ and
$\hat D$ are $d_T$ and $d_D$. The dominant hard scattering part multiplying
$\hat T$ and $\hat D$ then must behave as
\be
	\Gamma(p_i,k_j,q) \sim Q^{c-d_T-d_D},
\ee
where $c$ is a constant depending on the specific observable. The behaviour of
$\hat T$ is determined by its Lorentz structure and the specific component that
one considers. As in the matrix element every Lorentz-index contributes as
$P^\mu$
one finds after collecting the powers of $Q$ in the frame $A=Q$ where
$P^+ \sim Q$, $P^- \sim M^2 / Q$ and $P^\perp \sim M$
\be
	\hat T \sim M^{d_T-h_T} Q^{h_T}.
\ee
After determining $h_T$ in this way, the twist of the (nonlocal) operator in
$\hat T$ is defined as $t_T\equiv
d_T -h_T$. Similarly, the behaviour of $\hat D$, where $p_h^- \sim Q$,
$p_h^+ \sim m_h^2 / Q$ and $\pht \sim m_h$, is given by
\be
	\hat D \sim m_h^{d_D-h_D} Q^{h_D}
\ee
and the twist is defined as $t_D=d_D-h_D$. Note that we (following Jaffe)
discuss
twist for an operator which is in general nonlocal. Moreover, different
components
(defining different profile functions) will have different twist.
For example $\psib(x) \gamma^+ \psi(0)$  will have
$t=2$, but $\psib (x) \gamma^- \psi(0)$ has $t=4$. The twist for the $++\ldots
+$ component of a symmetric traceless operator is identical to the 'twist' as
used in the OPE approach for inclusive DIS. The name twist is still used in our
context as the above leads for  $O_{n,m}$ to the behaviour
 \be
	O(P,q,p_h) \sim \left( \frac{1}{Q} \right)^{t_T+t_D-c}.
\ee
The leading Born diagram starts off with  bilocal twist two operators in $\hat
T$
and $\hat D$, but also contains twist three and twist four pieces. Diagrams
with,
for instance, an extra pair of quark fields in $\hat T$ or $\hat D$ are
suppressed
by a factor $Q^{-2}$ as $t_T$ or $t_D$ is increased by 2.
For gluons the situation is more complicated, since addition of a gluon does
not
 change the twist. By choosing the gauge $A^+=0$ for the lower and $A^-=0$ for
the
upper blob we force the gluon contribution to add at least one unit of twist.
Let
it be clear however that one can add in general an arbitrary number
of longitudinally gluons to every diagram. Only after  a suitable choice of
gauge
 these contributions will vanish.

\subsection{\sf Correlation functions and profile functions}
 The profile functions will be defined as fourier transforms of projections of
$T$ and $D$. Explicit definitions of the profile functions are given in
appendix B.
 With these definitions we find
\beqa
	\int dp^- \ \mbox{Tr} \left[ T(p)\right]_{\pp=\xb \Pp}
	&=&
	\frac{q^-}{Q}\,\frac{2M\xb}{Q}\,e(\xb,{\mbox{\boldmath
	$p$}}_\perp),  \\
	\int dp^- \ \mbox{Tr} \left[T(p)\gamma^\mu\right]_{\pp=\xb\Pp}
	&=&
	\frac{q^-}{Q}
	\Biggl[ (\hat T^\mu - \hat q^\mu)   f^+(\xb, \bpt)  \nn \\
	&&\qquad\mbox{}+\frac{2\xb p_\perp^\mu}{Q }f^\perp(\xb,\bpt)  +
	{\cal O}\left(
	{1 \over Q^2} \right)\Biggr], \label{rel1}\\
	\int d\kp \ \mbox{Tr} \left[ D(k)\right]_{\km=p_h^- /z}
	&=&
	-\frac{q^+}{Q}\,\frac{2m_h}{Q}\,E(z,{\mbox{\boldmath
	$p$}}_{h\perp} - z{\mbox{\boldmath $k$}}_\perp ), \\
	\int d\kp \ \mbox{Tr} \left[ \gamma^\mu D(k)\right]_{\km=p_h^- /z}
	&=&
	 -\frac{q^+}{Q}
	\Biggl[ \left(\hat T^\mu+ \hat q^\mu\right)\,D^-(z, \pht-z\bkt) \nn \\
  	&&\qquad \mbox{} + \frac{2 (k_{\perp}^\mu-p_{h\perp}^\mu/z)}{zQ}\,
	D^\perp(z, \pht-z\bkt)
	\nn \\
	&&\qquad\mbox{}+\frac{2p_{h\perp}^\mu}{zQ}\, D^-(z, \pht-z\bkt)
	+  {\cal O}\left(
	{1 \over Q^2} \right)\Biggr], \label{rel2}
\eeqa
where $f^+$ and $D^-$ are twist two profile functions and $e$, $f^\perp$, $E$,
$D^\perp$ are twist three profile functions. These are all scalar functions.
The functions $f^+$ and $e$ are the same as defined by Jaffe and Ji in
\cite{JaffeJi}.
The function $D^-$ is the same as defined by Collins and Soper
in \cite{Collins}. The functions $f^\perp$ and $D^\perp$  are new.

 In order to handle the three-parton correlations $M$ and $F$ we define the
functions ${\cal M}$ and ${\cal F}$,  completely analogous to $M$ and $F$ but
with the replacement $A^\alpha \rightarrow
D^\alpha=\partial^\alpha-igA^\alpha$.
With partial integration we proof the relations
\beqa
	\int d^{\, 4}p_1 \ {\cal F}^\alpha_{ij}(p,p_1)
	&=&
	-ip^\alpha T_{ij}(p)-ig\int d^{\,4}p_1\
	F_{ij}^\alpha(p,p_1), \nn \\
	\int d^{\, 4}k_1 \ {\cal M}^\alpha_{ij}(k,k_1)
	&=&
	-ik^\alpha D_{ij}(k)-ig\int d^{\,4}k_1\
	M_{ij}^\alpha(k,k_1).
	\label{partial}
\eeqa
As argued before, the three-parton correlation functions $F$ and $M$ or
${\cal F}$ and ${\cal M}$ will be twist 3
in a suitably chosen  gauge. We will now show that the associated leading
 profile function is in fact expressible in the twist three profile functions
appearing in eq. (\ref{rel1}) and eq. (\ref{rel2}). For
this we use the QCD equations of motion \cite{JaffeJi,EFP,Poll}. We define
$\psi_{\pm} = P^{\pm}\psi$ where $P^{\pm}=\gamma^{\mp}\gamma^{\pm}/2$. It is
then straightforward to show that $\psi_+$ and $\mbox{\boldmath $A$}^\perp$
are dynamically independent fields in the gauge $A^+=0$ \cite{Kogut}. This
means that parts of the three-parton  correlation functions corresponding to
$\psi_-$ and $A^-$ can be written as correlation functions involving more than
three
partons and are thereby guaranteed to contribute at twist $>$ 3. From this
simple
 argument we know beforehand, that if we are interested in twist 3
contributions
(and we are!) then we only have to consider
$\mbox{Tr}[\gamma^+ F^\alpha (p,p_1)]$, with $\alpha=1,2$. For the $M$
functions
we find twist 3 for $\mbox{Tr}[\gamma^- M^\alpha (k,k_1)]$ in the gauge
$A^-=0$.
One draws the same conclusion from the twist analysis of the previous section.
As $\psib \gamma^+ \psi=\sqrt{2} \psi_+^\dagger \psi_+$ we conclude that
the leading contribution (in twist 2) comes from $\psi_+^\dagger A^+ \psi_+$.
When choosing $A^+=0$ the leading contribution is then at twist 3 and given by
$\psi_+^\dagger A^\perp \psi_+$.

In the $A^\pm=0$ gauge we deduce the equations of motion
\be
	D^\alpha(x)
 \gamma^{\pm}\psi(x)=\partial^\pm \gamma^\alpha\psi(x).
\ee
This is only true when applied between unpolarized nucleon states.
Applying these equations we find for the relevant transverse components
($\alpha=1,2$)
\beqa
	&&\frac{1}{2(2\pi)^4}\int dp^- d^{\, 4}x\ e^{-ip\cdot x}
	\bra P |\psib (x) \gamma^+ i D^\alpha (0) \psi (0) |P\ket
	=
	\xb \bpt^\alpha f^\perp ,\\
	&&\frac{1}{4z(2\pi)^4}\int d\kp d^{\, 4}x \ e^{ik'\cdot x}
	\mbox{Tr} \left.\bra 0| \psi (x) i D^\alpha(0)\ahk a_h \psib(0)
	\gamma^-|0\ket\right|_{\pht=0}
	=
	\frac{\bkt^{\prime\alpha}}{z}D^\perp.
\eeqa
Here  $k'$ is the quark momentum $k$ obtained after a   Lorentz transformation
to the frame where $\pht=0$,  $\bkt'=\bkt-\pht/z$   as explained in  appendix
B.
We have to transform back to the original frame with $\pht \neq 0$. Note that
because of this transformation
\be
	\frac{1}{4z(2\pi)^4}\int d\kp d^{\, 4}x \ e^{ik'\cdot x}
	\mbox{Tr} \left.\bra 0| \psi (x) i D^-(0)\ahk a_h \psib(0)
	\gamma^-|0\ket\right|_{\pht=0}=k^-D^-
\ee
will contribute to the transverse three-parton piece.
This gives us as the result in leading order which we will use in the next
section
\beqa
	\int dp^- \int d^{\, 4} p_1\ \mbox{Tr}\left[i {\cal F}^\alpha
	\gamma^\mu\right]_{\pp=-\qp}
	&=&
	\left(\frac{\qm}{Q}\right)\left[ (\hat T^\mu -\hat q^\mu )
	p_\perp^\alpha \ \xb f^\perp +
	{\cal O}\left(\frac{1}{Q^2}\right)\right], \label{rel3}\\
	\int d\kp \int d^{\, 4} k_1\ \mbox{Tr}\left[i {\cal M}^\alpha
	\gamma^\mu\right]_{\km=\qm}
	&=&
	\left(\frac{-\qp}{Q}\right)\left[ (\hat T^\mu +\hat q^\mu )
	\left[ k_\perp^{  \prime\alpha}\ {1 \over z}D^\perp+
	\frac{p_{h\perp}^{\alpha}}{z}D^-\right]
	+{\cal O}\left(\frac{1}{Q^2}\right)\right]. \label{rel4}
\eeqa

\section{\sf Calculation of the hadronic tensor and gauge invariance}

We present the calculation of the hadronic tensor $\cw$ from the  diagrams
of figures \ref{born} and \ref{gauge}. It consists  of  three parts,
$\cw=\cw_B+
\cw_F + \cw_M$, where $\cw_B$ is given by the Born diagram of figure
\ref{born},
$\cw_F$ by the one-gluon diagrams \ref{gauge}a and \ref{gauge}b and $\cw_M$ by
figures \ref{gauge}c and \ref{gauge}d.
\beqa
	\cw_B^{\mu\nu} &=& \frac{4z}{M} \int d^{\, 4}p\, d^{\, 4}k
	\ \delta^4(p+q-k) \mbox{Tr}\left[\gamma^\mu T(p)\gamma^\nu
	D(k)\right], \\
	 \cw_F^{\mu\nu} &=& \frac{-4z}{M} \int d^{\, 4}p \,d^{\, 4}k
	\ \delta^4(p+q-k) \int d^{\, 4}p_1\ \Bigl\lgroup \mbox{Tr}
	\bigl[ \gamma^\alpha \frac{(\slash p_1 +\slash q)}{(p_1 +q)^2}
	\gamma^\mu g F_\alpha(p,p_1)\gamma^\nu D(k)\bigr] \nn \\
	&& +\mbox{Tr} \bigl[\gamma^\nu \frac{(\slash p_1 +\slash q)}
	{(p_1 +q)^2}\gamma^\alpha D(k) \gamma^\mu g F_\alpha (p_1,p)
	\bigr]\Bigr\rgroup, \\
	 \cw_M^{\mu\nu} &=& \frac{-4z}{M} \int d^{\, 4}p \,d^{\, 4}k
	\ \delta^4(p+q-k) \int d^{\, 4}k_1\ \Bigl\lgroup \mbox{Tr}
	\bigl[ \gamma^\mu \frac{(\slash k_1 -\slash q)}{(k_1 -q)^2}
	\gamma^\alpha T(p)  \gamma^\nu gM_\alpha (k,k_1)\bigr] \nn \\
	&& +\mbox{Tr} \bigl[\gamma^\alpha \frac{(\slash k_1 -\slash q)}
	{(k_1 -q)^2}\gamma^\nu gM_\alpha (k_1,k) \gamma^\mu  T(p)
	\bigr]\Bigr\rgroup.
\eeqa
It is not necessary to consider the colour charge operators at this point, as
they will disappear again when the gluon fields are eliminated using the QCD
equations of motion.
As discussed in the introduction we will show that $\cw_F$ and $\cw_M$ generate
the first order of an expansion in $g$ of the linking exponential, needed to
render the Born term gauge invariant. We will discuss only the $F$ part as
the $M$ part is completely analogous.  The required term for gauge invariance
is
\beqa
	\cw^{\mu\nu}_{\mbox{\scriptsize Link},F}
	&=&
	\frac{4z}{M} \int d^{\, 4}p \,d^{\, 4}k
	\ \delta^4(p+q-k) \mbox{Tr} \Bigl[ \gamma^\mu_{kl}\,
	\frac{ig}{2(2\pi)^4} \int d^{\,4}x\ e^{-ip\cdot x} \nn \\
	&&\times \int d^{\, 4} y
	\bra P | \psib_i (x)\, \xi^\alpha_{k'k,jj'}(x-y,-y) A_\alpha(y)\,
	\psi_l(0)|P\ket \gamma^\nu_{ij} \nn \\
	&& \times\frac{1}{4z(2\pi)^4} \int d^{\, 4}
	z \ e^{i k\cdot z} \bra 0 | \psi_{j'} (z) \ahk a_h \psib_{k'}(0)
	|0\ket\Bigr],
\eeqa
where we have the condition on $\xi^\alpha$
\be
	\frac{\partial}{\partial y^\alpha}\xi^\alpha_{ k'k,jj'}(x-y,-y) =
	\left[ \delta^4 (y) - \delta^4 (x-y) \right] \delta_{jj'}\delta_{k'k}.
	\label{condi}
\ee
By comparing $\cw^{\mu\nu}_F$ with $\cw^{\mu\nu}_{\mbox{\scriptsize Link},F}$
we identify
\be
	\xi^\alpha_{ k'k,jj'}(x-y,-y) =
	\delta_{jj'} \int \frac{d^{\, 4}p_1}{(2\pi)^4} \left(
	\gamma^\alpha \frac{i}{\slash p_1 + \slash q}\right)_{k'k}
	e^{i(p-p_1)\cdot y} +
	\delta_{k'k} \int \frac{d^{\, 4}p_1}{(2\pi)^4} \left(
	 \frac{i}{\slash p_1 + \slash q}\gamma^\alpha\right)_{jj'}
	e^{i(p-p_1)\cdot (x-y)}.
	\label{ksi}
\ee
We calculate the derivative of $\xi^\alpha$ and use the identities
\beqa
	(p-p_1)^\alpha \frac{1}{\slash p_1 + \slash q}
	\gamma_\alpha &=& \frac{1}{\slash p_1 + \slash q} \slash k -1, \\
	(p-p_1)^\alpha \gamma_\alpha \frac{1}{\slash p_1 + \slash q}
	&=& \slash k \frac{1}{\slash p_1 + \slash q} -1,
\eeqa
which are essentially Ward-Takahashi identities for the hard scattering vertex.
In leading order we use the kinematics of eqs. (\ref{kin1}) and (\ref{kin2}) to
find
\be
	\frac{\partial}{\partial y^\alpha}\xi^\alpha_{ k'k,jj'}(x-y,-y) =
	\delta_{jj'}P^+_{k'k}\delta^4 (y) - \delta_{k'k}
	P^-_{jj'}\delta^4 (x-y)  + {\cal O}\left(\frac{1}{Q}\right).
\ee
Since for the $D$ correlation function the leading parts are given
by $\psi_-=P^- \psi$ fields in the matrix elements and $(P^-)^2=P^-$, we see
that in leading order our $\xi^\alpha$ fulfills condition (\ref{condi}). From
this  and the analogous reasoning for the other two diagrams we conclude that
the four diagrams of figure 4 are enough to render our calculation gauge
invariant
up to ${\cal O}(Q^{-2})$. Note that both in our derivation of the linking
operator
in the introduction as in our determination of the leading diagrams we have not
considered the contributions of  multiple longitudinal gluons. However, we have
proven gauge invariance for the ${\cal O}(g)$ term which in the gauge $A^+=0$,
$A^-=0$ respectively will be the leading term. In our calculations we will also
not need to consider these diagrams as we will choose these gauges.

Now that we have convinced ourselves that our starting expression is gauge
invariant
 we return to the calculation of the diagrams. We will expand
the correlation functions in the standard basis of Dirac matrices. Because of
time
reversal and parity invariance together with hermiticity only $1$ and
$\gamma^\mu$
can contribute in unpolarized electromagnetic scattering. Furthermore we can
proof
 with time reversal invariance and hermiticity that
$\mbox{Tr}\left[ \gamma^\mu F_\alpha (p,p_1) \right]$$=$$\mbox{Tr}
\left[\gamma^\mu F_\alpha(p_1,p)\right]$ and the same for $M_\alpha(k,k_1)$.
The tensors reduce to
\beqa
	\cw_B^{\mu\nu} &=& \frac{4z}{M} \int d^{\, 4}p\, d^{\, 4}k
	\ \delta^4(p+q-k) \frac{1}{16} \mbox{Tr}\left[ \gamma_\sigma T
	\right] \mbox{Tr}\left[\gamma_\lambda D\right]
	S_B^{\mu\sigma\nu\lambda}, \\
	\cw_F^{\mu\nu} &=& \frac{-4z}{M} \int d^{\, 4}p \,d^{\, 4}k
	\ \delta^4(p+q-k) \nn \\
	&&\int d^{\, 4}p_1\ \frac{(p_1+q)_\rho}{(p_1 +q)^2}
	\left\lgroup \frac{g}{16} \mbox{Tr} \left[\gamma_\sigma F_\alpha
	(p,p_1)\right] \mbox{Tr}\left[\gamma_\lambda D\right]\left\{
	S^{\alpha\rho\mu\sigma\nu\lambda}+
	S^{\nu\rho\alpha\lambda\mu\sigma} \right\}\right\rgroup, \\
 	 \cw_M^{\mu\nu} &=& \frac{-4z}{M} \int d^{\, 4}p \,d^{\, 4}k
	\ \delta^4(p+q-k) \nn \\
	&&\int d^{\, 4}k_1\  \frac{(k_1-q)_\rho}{(k_1 -q)^2}
	\left\lgroup \frac{g}{16} \mbox{Tr} \left[\gamma_\sigma  T
	\right] \mbox{Tr}\left[\gamma_\lambda M_\alpha(k,k_1)\right]\left\{
	S^{\mu\rho\alpha\sigma\nu\lambda}+
	S^{\alpha\rho\nu\lambda\mu\sigma} \right\}\right\rgroup,
\eeqa
where
\beqa
	S_B^{\mu\sigma\nu\lambda}&=&\mbox{Tr}\left[\gamma^\mu\gamma^\sigma
	\gamma^\nu\gamma^\lambda\right],\\
	S^{\alpha\rho\mu\sigma\nu\lambda} &=& \mbox{Tr}\left[
	\gamma^\alpha\gamma^\rho
	\gamma^\mu\gamma^\sigma\gamma^\nu\gamma^\lambda\right].
\eeqa
Contributions of the Dirac unit matrix either give rise to an odd number of
gamma matrices or are suppressed by $Q^{-1}$. Since the fermion propagator in
the frame $A=Q$ goes as $Q^{-1}$ we only have to pick out the order 1
contribution
of the rest of the integrand. The leading contribution is given by
\be
 	S^{\perp\rho\mu -\nu +}+S^{\nu\rho\perp +\mu -}
\ee
for $	\cw_F^{\mu\nu}$ and
\be
	S^{\mu\rho\perp -\nu +}+S^{\perp \rho \nu + \mu -}
\ee
for $ \cw_M^{\mu\nu}$.
Because $\gamma^+\gamma^+=\gamma^-\gamma^-=0$ and
$\{\gamma^\perp,\gamma^\pm\}=0$
this means that $\rho=-$ in the first case and $\rho=+$ in the second. This in
turn
implies that the fermion propagators are
in fact constants in leading order as can be read off from eq. (\ref{kin2}) in
the
frame $A=Q$
\beqa
 	\frac{(p_1+q)^\rho}{(p_1 +q)^2}&=&\frac{1}{2Q }\left(\hat T
	-\hat q\right)^\rho, \label{rel5} \\
	\frac{(k_1-q)^\rho}{(k_1 -q)^2}&=& -\frac{1}{2Q }\left(\hat T+
	\hat q\right)^\rho. \label{rel6}
\eeqa
Now we can take the propagators outside the $k_1$-integration and use the
relations (\ref{partial}). Furthermore we decompose $\delta^4(p+q-k)$ as
$\delta(\pp+\qp)\,\delta(\qm-\km)\,\delta^2(\bpt-\bkt)$. We finally find
\beqa
	\cw_B^{\mu\nu} &=& \frac{z}{4M} \int  dp^+ d\km d^{\, 2}\bpt\
	  \frac{1}{16} \mbox{Tr}\left[ \gamma_\sigma T
	\right] \mbox{Tr}\left[\gamma_\lambda D\right]
	S_B^{\mu\sigma\nu\lambda}, \\
	\cw_F^{\mu\nu} &=& \frac{-z}{4M} \int dp^+ d\km d^{\, 2}\bpt\
	\  \frac{(p_1+q)^\rho}{(p_1 +q)^2}\Bigg\{
	\Big\lgroup\int d^{\, 4}p_1\ \mbox{Tr}\left[
	i{\cal F}_\alpha\gamma_\sigma\right]\ \mbox{Tr}\left[\gamma_\lambda
	D\right] \nn \\
	&& \mbox{}-\mbox{Tr}\left[\gamma_\sigma T\right]\ \mbox{Tr}
	\left[\gamma_\lambda D\right] p_{\perp\alpha}\Big\rgroup
	\Big\lgroup S^{\alpha\rho\mu\sigma\nu\lambda}+
	S^{\nu\rho\alpha\lambda\mu\sigma}\Big\rgroup\Bigg\} , \\
	\cw_M^{\mu\nu} &=& \frac{-z}{4M} \int dp^+ d\km d^{\, 2}\bpt\
	\  \frac{(k_1-q)^\rho}{(k_1-q)^2}\Bigg\{
	\Big\lgroup \mbox{Tr}\left[
	 T\gamma_\sigma\right]\ \int d^{\, 4}k_1\mbox{Tr}\left[
	i{\cal M}_\alpha \gamma_\lambda
	\right] \nn \\
	&& \mbox{}-\mbox{Tr}\left[\gamma_\sigma T\right]\ \mbox{Tr}
	\left[\gamma_\lambda D\right] k_{\perp\alpha}\Big\rgroup
	 \Big\lgroup S^{\mu\rho\alpha\sigma\nu\lambda}+
	S^{\alpha\rho\nu\lambda\mu\sigma} \Big\rgroup\Bigg\}.
\eeqa
This is the contribution of the  relevant diagrams, expressed in quantities
which we have given as tensors in next-to-leading order on the basis $\{\hat
T$,
$\hat q$, $p_\perp$,$k_\perp$,$p_{h\perp}\}$ in the previous section. To be
precise, we will substitute the relations (\ref{rel1}), (\ref{rel2}),
(\ref{rel3}), (\ref{rel4}), (\ref{rel5}) and (\ref{rel6}). Note that
${\hat T}^2=1$, ${\hat q}^2=-1$, $\hat T \cdot \hat q=0$ and that they do
not have transverse components. The calculation is then very well suited for
symbolic manipulation with a computer. We have used    FORM \cite{form} to
evaluate the total hadronic tensor
as
\beqa
	\widetilde \cw^{\mu\nu}(P,q,p_h,\bpt) &=&
	\left( -{\hat q}^\mu {\hat q}^\nu - g^{\mu\nu}\right)\left(
	\frac{z}{M} f^+ D^- \right)\nn\\
	&&\mbox{}+\left({\hat T}^\mu {\hat T}^\nu \right)\left(
	\frac{z}{M} f^+ D^- \right)\nn\\
	&&\mbox{}+\left({\hat T}^\mu p_{\perp}^\nu +
	 {\hat T}^\nu p_{\perp}^\mu\right)
	\left( \frac{2xz}{MQ}f^\perp D^-+ \frac{2}{MQ} f^+ D^\perp -
	\frac{2z}{MQ} f^+ D^- \right) \nn\\
	&&\mbox{} + \left({\hat T}^\mu p_{h\perp}^\nu +
	 {\hat T}^\nu p_{h\perp}^\mu\right)
	\left(\frac{2}{MQ} f^+ D^- -\frac{2}{zQM} f^+ D^\perp\right)
	+{\cal O}\left( \frac{1}{Q^2} \right),
\eeqa
where
\be
\cw^{\mu\nu} = \int d^{\, 2}\bpt \widetilde \cw^{\mu\nu}.
\ee
One can convince oneself that $q_\mu \widetilde \cw^{\mu\nu} ={\cal
O}(Q^{-2})$.
The hadronic tensor $\widetilde \cw^{\mu\nu}$ is in fact the hadronic tensor
for
the semi-inclusive  process where one observes  in addition to a hadron $h$ the
transverse momentum of the jet, i.e. the transverse momentum of the ejected
quark
in our approach
\be
	\frac{d\sigma^{(e+H\rightarrow e' + h + \mbox{\scriptsize jet})}}
	{d^{\, 3}p_h d^{\, 2}\bp_{\perp} d\Omega dE' } = \frac{\alpha^2}{Q^4}
	\frac{E'}{E} \frac{1}{2E_h} L_{\mu\nu} \widetilde \cw^{\mu\nu},
\ee
 As mentioned in section 2 this gives then rise to two extra structure
functions.
 We project out the leading structure functions (all other structure functions
are
of order $Q^{-2}$),
\beqa
	\widetilde \cw_T(\xb,z,\pht, \bp_{\perp}) &=& \frac{z}{M}  \:
 	f^+(\xb,\bp_{\perp}) D^-(z, \pht-z\bp_{\perp}) +
	{\cal O}(\frac{1}{Q^2}),  \nn\\
	\widetilde \cw_{LT}^1(\xb,z,\pht, \bp_{\perp}) &=&
	\frac{|\bp_{h\perp}|}{Q}\frac{4}{M}\Big[
	f^+(\xb,\bpt) \nn \\
	&& \times \big\lgroup\frac{1}{z}D^\perp (z,\pht-z\bpt)-
	D^-(z,\pht-z\bpt)\big\rgroup \Big],	\nn\\
	\widetilde \cw_{LT}^2(\xb,z,\pht, \bp_{\perp}) &=&
	\frac{|\bpt|}{Q}\frac{4}{M}\Big[
	z\big\lgroup f^+(\xb,\bpt)-\xb f^\perp (\xb,\bpt) \big\rgroup
	D^-(z,\pht-z\bpt) \nn \\
	&&\mbox{}- f^+(\xb,\bpt)D^\perp (z,\pht-z\bpt)
	\Big].
	\label{struc}
\eeqa

\section{\sf Results}
\subsection{\sf The quark distribution and fragmentation function}
In this last part we will present the cross sections of several processes
in lepton-hadron scattering. These cross sections will be expressed in the
previously derived quark profile functions. Firstly we will examine the
properties of these functions.

The profile function $f^+$ is just the well known quark distribution function,
 see \cite{JaffeAlamos}. The function
\beqa
	&&f_{q/H}(\xb, \bp_{\perp} )=f^+(\xb,\bpt) \nn \\
	&&\qquad=\frac{1}{2(2\pi)^3 }
	\int dx^-\,d^{\, 2}{\mbox{\boldmath $x$}}_\perp
	e^{i(q^+x^- +\, {\mbox{\boldmath $p$}}_\perp \cdot {\mbox{\boldmath
	$x$}}_\perp)} \
	\left.\bra P \vert   \psib(x) \gamma^+ \psi(0) \vert P \ket
	\right|_{x^+ = 0},
\eeqa
is interpreted as the probability  of finding a quark
with lightcone momentum fraction $\xb = \pp / P^+=-\qp/ \Pp$ and transverse
momentum
$\bp_{\perp}$ in a target with no transverse momentum. This can be seen when
one
quantizes the quark fields on the lightcone with $\xp=0$ (see \cite{JaffeJi}).
As the $\psi_+$ fields
are the essentially free fields one can then substitute a free field expansion.
The operator in the matrix element then essentially reduces to a quark number
operator, counting quarks with momentum fraction $\xb$ and transverse momentum
$\bpt$. The integral over $\xb$ and $\bpt$ comes out as 1, so $f^+$ is a
probability distribution. Upon integration over the transverse momentum one
finds the
the quark distribution
\be
	f_{q/H}(\xb )=\frac{1}{4\pi }
	\int dx^-\,
	e^{i q^+x^- } \
	\left.\bra P \vert \psib(x) \gamma^+ \psi(0) \vert P \ket
	\right|_{x^+={\mbox{\boldmath $x$}}_\perp = 0}.
\ee
With this form one can easily prove sum rules expressing probability and
momentum
conservation at the quark level. As a check on the answer one can
consider a quark target. One expects that the probability of finding a quark
in a quark reduces to a delta function in $\xb$ and $ {\mbox{\boldmath
$p$}}_\perp$.
 An explicit calculation, substituting a free quark state
for the target state $|P\ket$ leads to
\be
	f_{q/q} (\xb, \bp_{\perp} )= \delta (1-\xb) \delta^2 (\bp_{\perp}).
\ee

The fragmentation function
\beqa
	&&D_{h/q}(z,\pht)=D^-(z,\pht) \nn \\
	&&\qquad =\frac{1}{4z(2\pi)^3}  \int dx^+\,d^{\,2}{\mbox{\boldmath $x$}}_\perp
	e^{i(q^-x^+  +\frac{\pht \cdot  {\mbox{\boldmath $x$}}_\perp }{z})}
	\left.
	\mbox{Tr}\,\bra 0 \vert \psi(x) \,a_{h}^\dagger a_{h}\,\psib(0)
	\gamma^- \vert 0 \ket  \right|_{x^- = 0, \pht={\bf 0}}, \nn \\
\eeqa
is interpreted as the multiplicity of hadrons (of a certain type) found
in the hadronization products (jet) of a quark with transverse momentum zero,
hadrons with momentum fraction $z=p^-_h/\km=p^-_h/ \qm$ and transverse momentum
$\pht$.
Analogous to the case of the distribution function this can be seen  by
quantizing
the quark fields on the lightcone, but now for $\xm=0$. One then finds
essentially
a hadron number operator acting between two quark states. The sumrules
discussed hereafter then confirm that $D^-$ is a  multiplicity  distribution.
If the original quark does  have a transverse momentum, one should make the
substitution $\pht \rightarrow \pht'=\pht-z\bk_\perp$, where $\pht'$ is then
the transverse momentum of the hadron with respect to the quark with momentum
$\bk_\perp$. Upon integration over $\pht$ one finds
\be
	D_{h/q}(z)=
	\frac{z}{8\pi}  \int dx^+\,e^{iq^-x^+  } \left.
	\mbox{Tr}\,\bra 0 \vert \psi(x) \,a_{h}^\dagger a_{h}\,\psib(0)
	\gamma^- \vert 0 \ket  \right|_{x^-={\mbox{\boldmath $x$}}_\perp = 0,
	\pht={\bf 0}}.
\ee
This is the scaling fragmentation function. Both forms were first defined in
\cite{Collins} and also used in \cite{psu102}. By integrating over $z$ and
using the techniques of appendix
A we can derive the sum rule
\be
\int dz\:D_{h/q}(z)=\langle n_h \rangle_q,
\ee
expressing the fact that $D(z)$ is a multiplicity distribution. In
\cite{Collins}
it was already proven that
\be
\sum_{\mbox{\scriptsize hadrons}}\int dz \: zD(z) =1,
\ee
which expresses momentum conservation. Finally we can also here consider the
production of a quark from a quark. We substitute free quark operators for the
$\ahk, \ah$ and find
\be
D_{q/q} (z, \bp_{q\perp}) = \delta(1-z)\delta^2(\bp_{q\perp}).
\ee

The profile functions $f^\perp$ and $D^\perp$ do not have an interpretation
as a distribution function, since they depend on interacting quark fields so
no free field expansion can be applied.
 The anti-quark contributions can be derived completely analogously
and lead to
\beqa
f_{\overline{q}/H}(\xb, \bp_{\perp} )&=&\frac{1}{2(2\pi)^3 }
\int dx^-\,d^{\, 2}{\mbox{\boldmath $x$}}_\perp
e^{i(q^+x^- +\, {\mbox{\boldmath $p$}}_\perp \cdot {\mbox{\boldmath
$x$}}_\perp)} \
\left.\mbox{Tr}\bra P \vert  \psi(x)  \psib(0)\gamma^+ \vert P \ket
\right|_{x^+ = 0},  \\
D_{h/\overline{q}}(z,\pht)&=&
\frac{1}{4z(2\pi)^3}  \int dx^+\,d^{\,2}{\mbox{\boldmath $x$}}_\perp
e^{i(q^-x^+  +\frac{\pht \cdot  {\mbox{\boldmath $x$}}_\perp }{z})} \left.
 \bra 0 \vert \psib(x)\gamma^- \,a_{h}^\dagger a_{h} \psi(0)
 \vert 0 \ket  \right|_{x^- = 0, \pht={\bf 0}}. \nn \\
&&
\eeqa
Inclusion of different flavours gives rise to a sum over contributing flavours
and quark charge
factors $e_q$.
\subsection{\sf Semi-inclusive cross-sections}

 Starting with the hadronic tensor derived in section 4, several
cross sections for semi-inclusive lepton hadron scattering can be expressed in
distribution and fragmentation functions. We will present a
cross section for the process $e+H\rightarrow e' + h + \mbox{jet}$, i.e. not
only a hadron is detected, but also the axis of the jet it belongs to. Other
cross sections will be determined from this one by integrating over measured
variables. With the structure functions (\ref{struc}) and accounting for
antiquarks
and flavour we have
\beqa
	\frac{d\sigma^{(e+H\rightarrow e' + h + \mbox{\scriptsize jet})}}
	{ d\xb dydzd^{\,2}\!\!\pht d^{\, 2} \bp_{\perp}}&=&
	\frac{8\pi\alpha^2ME}{Q^4}\sum_{i = q,\overline{ q}}
	e_i^{2} \Bigg\{\left\lgroup
	\frac{y^2}{2}+1-y\right\rgroup\xb f_{i/H}(\xb, \bp_{\perp})
	D_{h/i}(z, \pht-z\bp_{\perp}) \nn \\
	&&\mbox{}+2(2-y)\sqrt{1-y}\cos\phi_h\frac{|\pht|}{Q}\frac{\xb}{z}
	\Big[ f_{i/H}(\xb, \bp_{\perp}) \nn \\
	&&\times \left\lgroup\frac{1}{z}
	D^\perp_{h/i}
	(z,\pht-z\bpt)-D_{h/i}(z,\pht-z\bpt)\right\rgroup \Big] \nn \\
	&&\mbox{}+2(2-y)\sqrt{1-y}\cos\phi_j\frac{|\bpt|}{Q}\frac{\xb}{z}
	\Bigg[ 	- f_{i/H}(\xb,\bpt)D^\perp_{h/i} (z,\pht-z\bpt) \nn \\
	&& \mbox{}+z\left\lgroup f_{i/H}(\xb,\bpt)-\xb f^\perp_{i/h}
	 (\xb,\bpt)\right\rgroup
	D_{h/i}(z,\pht-z\bpt)	\Bigg]\Bigg\}.
\eeqa
Here $\phi_h$ and $\phi_j$ are the azimuthal angles of hadron and jet
with respect to the virtual photon direction. Note that $\pht$ is defined with
respect to the z-axis defined by the virtual
photon, but the transverse momentum entering the fragmentation function $\pht'=
\pht-z\bp_{\perp}$ is defined with respect to the jet axis.

In our  approach the transverse momentum of the jet immediately reflects the
transverse momentum of the quarks within the target.
To determine the jet cross section from this one has to first integrate over
$\pht$.
 A direct integration over
$\pht$ yields
\beqa
	\frac{d\sigma^{(e+H\rightarrow e' + h + \mbox{\scriptsize jet})}}
	{ d\xb dydz  d^{\, 2} \bp_{\perp}}&=&
	\frac{8\pi\alpha^2ME}{Q^4}\sum_{i = q,\overline{ q}}
	e_i^{2} \Bigg\{\left\lgroup
	\frac{y^2}{2}+1-y\right\rgroup\xb f_{i/H}(\xb, \bp_{\perp})
	D_{h/i}(z)\nn \\
	&&\mbox{}-2(2-y)\sqrt{1-y}\cos\phi_j\frac{|\bpt|}{Q}    \xb^2
	f^\perp_{i/H}(\xb,\bpt)D_{h/i} (z)  \Bigg\}
	\label{jet}
\eeqa
which gives for the jet cross section,
\beqa
\frac{d\sigma^{(e+H\rightarrow e' + h + \mbox{\scriptsize jet})}}
	{ d\xb dy d^{\, 2} \bp_{\perp}}&=&
	\frac{8\pi\alpha^2ME}{Q^4}\sum_{i = q,\overline{ q}}
	e_i^{2} \Bigg\{\left\lgroup
	\frac{y^2}{2}+1-y\right\rgroup\xb f_{i/H}(\xb, \bp_{\perp})\nn \\
	&&\mbox{}-2(2-y)\sqrt{1-y}\cos\phi_j\frac{|\bpt|}{Q}    \xb^2
	f^\perp_{i/h}(\xb,\bpt)   \Bigg\}
\eeqa
Note that we pick up a factor
$\langle n_h \rangle$ on both sides of the equation when integrating over $z$,
which cancels. With this cross section we can calculate an expectation value
for
the azimuthal angle $\phi_{\mbox{\scriptsize jet}}$,
\be
	\langle \cos \phi_{\mbox{\scriptsize jet}} \rangle=
	- \left(\frac{2|\bp_\perp |}{Q} \right)\frac{(2-y)
	\sqrt{1-y}}{2(1-y)+y^2}\left(\frac{\xb f^\perp(\xb,\bpt)}{f(\xb,\bpt)}
	\right),
\ee
where the structure factor given by the ratio of $\xb f^\perp$ and $f$ comes as
an extension of the result previously derived by  Cahn in \cite{Cahn} for the
case of free quarks.    Integrating over $\bp_\perp$ in eq. (\ref{jet})
we find the one-particle semi-inclusive cross section. The $z$ dependence is
simply given by a factor $D(z)$, and the cosine integrates to zero due to the
azimuthal symmetry of $f^\perp(\xb, \bp_\perp)$,
\be
\frac{d\sigma^{(e+H \rightarrow e'+h)}}{d\xb dy dz}=
\frac{8\pi\alpha^2 ME}{Q^4} \left[ (1-y) + \frac{y^2}{2} \right]\sum_{i = q,
\overline{ q}} e_i^{2}
 \xb
f_{i/H}(\xb) D_{h/i}(z).
\ee
If we compare this with eq. (\ref{xyz}) we see that we have recovered the
factorization
and scaling of the semi-inclusive structure function
\be
H_2(\xb,Q^2,z)=H_2(\xb,z)=\sum_{q} e_{q}^2\, \xb\left[  f_{q/H}(\xb) D_{h/q}
(z)
+f_{\overline{q}/H}(\xb) D_{h/\overline{q}}(z)\right]
,
\ee
and the semi inclusive Callan-Gross relation
 \be
2\xb H_1 (\xb, z ) = H_2 (\xb, z ).
\ee
We recover the Callan-Gross relation because the corrections coming from the
twist 3 diagrams are only entering in $\ch_3$ and $\ch_4$.
Finally we will consider the cross section (\ref{bergerlang}) from which we
started for $l+H$$\rightarrow$$l'+h+X$. To this end we have to integrate over
the jet transverse momentum. For the functions $\ch_i$ occurring in eq.
(\ref{bergerlang})
 we find
\beqa
	\ch_1(\xb,z,Q^2,\pht^2)&=& \frac{1}{2}\sum_{i = q,\overline{ q}} e_i^{2}
	\int d^{\,2} \bp_\perp f_{i/H}(\xb,\bp_\perp )
	D_{h/i}(z,\pht-z\bp_\perp), \nn \\
	\ch_2(\xb,z,Q^2,\pht^2)&=&\xb\sum_{i=q,\overline{q}} e_{i}^{2}
	\int d^{\,2} \bp_\perp  f_{i/H}(\xb,\bp_\perp )
	D_{h/i}(z,\pht-z\bp_\perp), \nn \\
	\ch_3(\xb,z,Q^2,\pht^2)&=&\frac{2\xb}{z}\sum_{i=q,\overline{ q}}
	e_{i}^{2}
	\int d^{\,2} \bp_\perp \Bigg\{
	\frac{\bp_\perp\cdot\pht}{\pht^2}
	\Big[ 	- f_{i/h}(\xb,\bpt)D^\perp_{i/h} (z,\pht-z\bpt) \nn \\
	&& \mbox{}+z\left\lgroup f_{i/h}(\xb,\bpt)-\xb f^\perp_{i/h}
	 (\xb,\bpt)\right\rgroup
	D_{i/h}(z,\pht-z\bpt)	\Big] \nn \\
	&&\mbox{}+f_{i/H}(\xb, \bp_{\perp})
	 \left\lgroup\frac{1}{z}
	D^\perp_{h/i}
	(z,\pht-z\bpt)-D_{h/i}(z,\pht-z\bpt)\right\rgroup\Bigg\} \nn\\
\eeqa
To find the last expression we   used
\be
\int d^{\, 2} \bp_\perp\:p_{\perp}^{\mu}f(\xb, \bp_\perp )
D(z,\pht-z\bp_\perp )=\frac{p_h^\mu}{\pht^2}\int d^{\, 2} \bp_\perp\:
(\bp_\perp\cdot\pht)f(\xb, \bp_\perp )
D(z,\pht-z\bp_\perp ).
\ee
We conclude that these structure functions do not factorize without extra
assumptions on their transverse momentum dependence. However we do read off
 an extension of the Callan-Gross relation,
\be
 \ch_2 (\xb, z,\pht^2)=2\xb \ch_1 (\xb, z, \pht^2) .
\ee

\section{\sf Conclusions}

In this paper we have analyzed the structure functions in semi-inclusive
 deep-inelastic lepton-hadron scattering. We have considered all four
 structure functions, two of which ($\cw_{LT}$ and $\cw_{TT}$) can
only be measured by considering explicitly the momentum component of
the produced hadron perpendicular to the virtual photon momentum, or
 by considering explicitly
the perpendicular momentum of the produced jet. In all cases we have
only considered the contribution of one (forward or current) jet being
 produced
and particles therein. In the analysis we have included the twist two
and twist three contributions.

The twist two pieces in the structure functions can be expressed in terms
 of the well-known quark distribution and fragmentation functions which
can be considered as specific projections of quark-quark correlation
 functions (profile functions). The twist three pieces, which
 constitute the main contributions in $\cw_{LT}$ and $\cw_{TT}$
involve two new profile functions that appear as projections of
quark-quark correlation functions. Up to order $1/Q$, however,
one must also include quark-quark-gluon correlation functions in
order to obtain a gauge invariant result. By virtue of the QCD equations
 of motion the contributing pieces do not introduce new profile functions.
 The inclusion of
the $1/Q$ contributions does not affect the evolution of the twist two
 profile functions (the quark distribution and fragmentation functions).
We have not considered the evolution of the new profile functions, nor
the ${\cal O}(\alpha_s /Q)$ contributions in the process. An extension
of the analysis to
${\cal O}(Q^{-2})$ along the same lines as discussed here seems to be
very hard to us. In that situation not only several new profile functions
 will appear but one has also lost the separation between kinematic
quantities appearing in the distribution region from those appearing
in the fragmentation region.

Results in the on-shell parton model such as those  previously derived by Cahn
\cite{Cahn}
follow immediately as a special case by defining the profile functions with
respect to a free quark target, i.e. the amplitudes in (\ref{ampl}) are then
${\cal A}_1={\cal A}_2=4\pi^4\delta^4 (P-p)$ and ${\cal B}= 8\pi^4 m_q
\delta^4(P-p)$. In this case the azimuthal asymmetry $\langle \cos \phi
\rangle$
is of purely kinematical origin.

The analysis in terms of quark correlation functions is of interest as it
provides
a convenient method to relate cross sections of hard processes directly to
matrix
elements of (nonlocal) quark and gluon operators. The latter could in principle
be
calculated when we knew how to solve QCD, or they can be calculated in a model.
Although some data for semi-inclusive muon scattering exist
\cite{exp} we will not discuss them here. Neither have we presented any model
as this would require extra assumptions. In the presented analysis a minimal
amount of assumptions was made, namely those needed for the factorization of
the hard scattering part. We also want to point out that
one can do exactly the same analysis for Drell-Yan processes and
 $e^+e^-$-annihilation, thereby testing the validity of this approach and
extending the experimental possibilities of measuring the profile
 functions.

We would like to thank J.W. Bos, M. Dekker, A. W. Schreiber and R.D. Tangerman
for stimulating discussions. This work is supported by the foundation for
Fundamental Research of Matter (FOM) and the National Organization for
Scientific Research (NWO).

\appendix
\section{\sf Intermediate states and completeness}
The cross sections we are considering, describe the detection of one hadron in
correlation with a scattered electron. In an experiment often many hadrons of
the same type are produced in one event. This means that we have to pay special
attention not only when we are integrating over the hadronic variables when we
want to find an inclusive result, but also when   we isolate an hadronic
operator.
In this appendix we discuss how to treat these situations, based
on an example situation with only one type of scalar particles. Extensions to
flavour and spin are obvious. An $n$-particle state is denoted as $|p_1, \ldots
,p_n \ket$ and we abbreviate the phase space integrations as
$\dtp = \ddp{p}{E_p}$.   The order of the particles appearing in the states has
no meaning. Therefore the identity in this   space looks like \cite{Itzykson}
\be
\ci =  \sum_{n=0}^{\infty} \ci_{n},
\ee
where
\be
\ci_{n}=\frac{1}{n!}\int \dtp_1 \cdots \dtp_n\:
a^{\dagger}(p_1 )\cdots a^{\dagger}(p_n )|0\ket\bra 0|a(p_1 )\cdots a(p_n )
\label{in}
\ee
and
\be
\ci_{0}=|0\ket\bra 0|.
\ee
We first consider the situation  with    a multi-particle state where
we integrate over all particles except one, the situation as we encounter it
in our expressions. We denote with $P_X$ this complete system minus one
particle,
\beqa
\int \ddp{P_X}{E_X} |P_X,p_h \ket\bra P_X,p_h|&=&
|p_h\ket\bra p_h|+\int \dtp_1|p_1,p_h\ket\bra p_1,p_h |\nn \\ &&+
\frac{1}{2!} \int \dtp_1\dtp_2 |p_1,p_2,p_h\ket \bra p_1,p_2,p_h|+\ldots \nn \\
&=& \ahk {\cal I} a_h =\ahk a_h
\eeqa
 thus we can replace the sum minus one particle by the hadronic number
operator.
Then we want to integrate over $p_h$ to reconstruct an inclusive formula.
\beqa
\int \ddp{p_h}{E_h} \ddp{P_X}{E_X} |P_X,p_h \ket\bra P_X,p_h|&=&
 \int \dtp_h |p_h\ket\bra p_h| +
\int \dtp_h \dtp_1 |p_1, p_h \ket\bra p_1,p_h|\nn \\
&&+\frac{1}{2!}
\int\dtp_h\dtp_1\dtp_2 |p_1,p_2,p_h\ket\bra p_1,p_2,p_h|+\ldots \nn \\
&&\mbox{\hspace{0.1cm}}
\eeqa
This is the identity operator, but with every term separately multiplied with
 the number of particles. The summation over all states then yields just
the average number of particles $h$ produced in the processes considered,
\be
\int \ddp{p_h}{E_h} \ddp{P_X}{E_X} |P_X,p_h \ket\bra P_X,p_h|=
 \sum_{n} n {\cal I}_n   .
\ee
 Evaluated between particle states and reexpressed as cross sections this
relation  reads
\be
\int d\Omega_h dE_h \frac{d\sigma}{d\Omega_h dE_h d\Omega_e dE_e} =
\langle n_h (\Omega_e,E_e) \rangle \frac{d\sigma}{ d\Omega_e dE_e}.
\ee

\section{\sf Structure of the correlation functions}

In this appendix we study the  projections of the correlation functions
defined as
\begin{eqnarray}
	f_A(P;\xb,{\mbox{\boldmath $p$}}_\perp)
	& = &
	\left.\frac{1}{2(2\pi)^3}\int dx^-\,d^2{\mbox{\boldmath $x$}}_\perp
	e^{i(q^+x^- +\, {\mbox{\boldmath $p$}}_\perp \cdot {\mbox{\boldmath
	$x$}}_\perp)} \
	\bra P \vert \psib(x) \Gamma_A \psi(0) \vert P \ket
	\right|_{x^+ = 0} \\
	4zD_B(p_h;z,{\mbox{\boldmath $k$}}_\perp)
	& = &
	\left. 	\frac{1}{(2\pi)^3}
	\int dx^+\,d^2{\mbox{\boldmath $x$}}_\perp
	e^{i(q^-x^+ -\, {\mbox{\boldmath $k$}}_\perp \cdot {\mbox{\boldmath
	$x$}}_\perp)} \
	\mbox{Tr}\,\bra 0 \vert \psi(x) a_h^\dagger a_h \psib(0)\,\Gamma_B
	\vert 0 \ket
	\right|_{x^- = 0}
\end{eqnarray}
(with $q^+$ = $-\xb P^+$ and $q^-$ = $p_h^-/z$).
Using parity and time reversal invariance as well as hermiticity one proofs
that the most general structure of the matrix elements is
\begin{eqnarray}
	\bra P \vert \psib_i(x) \psi_j(0) \vert P \ket
	& = &
	\int \frac{d^4p}{(2\pi)^4}\, e^{i\,p\cdot x} \,
	\left[{\cal A}_1(P,p) \slash p +
	{\cal A}_2(P,p) \slash P + {\cal B}(P,p) \right]_{ji}, \label{ampl}\\
	\bra 0 \vert \psi_i(x)\, a_h^\dagger a_h\, \psib_j(0) \vert 0 \ket
	& = &
	\int \frac{d^4k}{(2\pi)^4}\, e^{-i\,k\cdot x} \,
	\left[{\cal C}_1(p_h,k) \slash k +
	{\cal C}_2(p_h,k) \slash p_h+ {\cal D}(p_h,k) \right]_{ij}.
\end{eqnarray}
At this point one immediately sees that for unpolarized distributions
$f_\alpha$ and $D_\beta$ are zero except for $\Gamma_{A,B}$ = $\gamma^\mu$ or
$\Gamma_{A,B}$ = $1$.

Consider as the first case $\Gamma_A$ = $\gamma^\mu$ for the $f$-profile
functions.
 One  obtains (choosing a frame where ${\mbox{\boldmath $P$}}_\perp$ = 0)
\begin{eqnarray}
	f^{[\gamma^+]}(P;\xb,{\mbox{\boldmath $p$}}_\perp)
	& = &
	\left.
	\frac{1}{2(2\pi)^3 }
	\int dx^-\,d^2{\mbox{\boldmath $x$}}_\perp
	e^{i(q^+x^- +\, {\mbox{\boldmath $p$}}_\perp \cdot {\mbox{\boldmath
	$x$}}_\perp)} \
	\bra P \vert \psib(x) \gamma^+ \psi(0) \vert P \ket
	\right|_{x^+ = 0} \nonumber \\
	& = &
	\left. \frac{1}{2(2\pi)^3 } \int \frac{dp^-}{2\pi}\
	\mbox{Tr}\, \left[ \gamma^+ ({\cal A}_1 \slash p +
	{\cal A}_2 \slash P+ {\cal B}) \right]
	\right|_{p^+ = -q^+} \nonumber \\
	& = &
	\left.\frac{1}{(2\pi)^4 } \int dp^2\ \left\lgroup
	{\cal A}_1(P,p)+\frac{1}{\xb} {\cal A}_2(P,p)\right\rgroup
	\right|_{p^+ = -q^+}
	\equiv \ f^+(\xb,{\mbox{\boldmath $p$}}_\perp ), \\
	f^{[\gamma^-]}(P;\xb,{\mbox{\boldmath $p$}}_\perp)
	& = &
	\left.
	\frac{1}{2(2\pi)^3 }
	\int dx^-\,d^2{\mbox{\boldmath $x$}}_\perp
	e^{i(q^+x^- +\, {\mbox{\boldmath $p$}}_\perp \cdot {\mbox{\boldmath
	$x$}}_\perp)} \
	\bra P \vert \psib(x) \gamma^- \psi(0) \vert P \ket
	\right|_{x^+ = 0} \nonumber \\
	& = &
	 \left. \frac{1}{2(2\pi)^3 } \int \frac{dp^-}{2\pi}\
	\mbox{Tr}\, \left[ \gamma^- ({\cal A}_1 \slash p +
	{\cal A}_2 \slash P +{\cal B}) \right]
	\right|_{p^+ = -q^+} \nonumber \\
	& = &
	\left.\frac{M^2}{2(2\pi)^4 (q^+)^2} \int dp^2 \,\left\lgroup
	\frac{(p^2 + {\bf p}_\perp^2)}{M^2} {\cal A}_1(P,p)+
	\xb {\cal A}_2(P,p) \right\rgroup \right|_{p^+ = -q^+} \nn \\
	&  \equiv &
	\ \frac{M^2\xb^2}{2(q^+)^2}\,
	f^-(\xb,{\mbox{\boldmath $p$}}_\perp ), \\
	f^{ [\gamma_\perp]}(P;\xb,{\mbox{\boldmath $p$}}_\perp)
	& = &
	\left.
	\frac{1}{2(2\pi)^3 }
	\int dx^-\,d^2{\mbox{\boldmath $x$}}_\perp
	e^{i(q^+x^- +\, {\mbox{\boldmath $p$}}_\perp \cdot {\mbox{\boldmath
	$x$}}_\perp)} \
	\bra P \vert \psib(x) \mbox{\boldmath $\gamma_\perp$}
 	 \psi(0) \vert P \ket
	\right|_{x^+ = 0} \nonumber \\
 	&=&
	\left. \frac{1}{2(2\pi)^3 } \int \frac{dp^-}{2\pi}\
	\mbox{Tr}\, \left[ \mbox{\boldmath $\gamma_\perp$} ({\cal A}_1 \slash p
 	 +{\cal A}_2 \slash P +
 	{\cal B}) \right]
	\right|_{p^+ = -q^+} \nonumber \\
	& = &
	-\frac{{\mbox{\boldmath $p$}}_\perp}{(2\pi)^4 q^+} \int dp^2 \,
	{\cal A}_1(P,p)  \equiv -\frac{\bp_\perp \xb}{\qp}f^\perp
	(\xb,\bp_\perp).
\end{eqnarray}
Here $f^+,f^-$ and $f^\perp$ are the profile functions. We assume that the
functions ${\cal A}_i$ and ${\cal B}$ have no singularities, so that the
profile
functions are all of order 1. In the frame $A=Q$ we also see at which twist
these functions enter the calculations. When $f^+$, i.e. the normal parton
distribution function, enters at ${\cal O}(1)$, then $f^\perp$ enters at
${\cal O}(Q^{-1})$ and $f^-$ at ${\cal O}(Q^{-2})$, as we expected from
the discussion in section 3.1.
   It is straightforward to express $f^{[\gamma^\mu]}$ as
\be
	f^{[\gamma^\mu]}(P;\xb,{\mbox{\boldmath $p$}}_\perp)  = \frac{q^-}{Q}
	\left[ (\hat T^\mu - \hat q^\mu)   f^+
	+\frac{2\xb p_\perp^\mu}{Q }f^\perp + {\cal O}\left(
	{1 \over Q^2} \right)\right].
\ee

The function $f^{[1]}$ introduces another  profile function
$e(\xb,{\mbox{\boldmath $p$}}_\perp)$,
\begin{eqnarray}
	f^{[1]}(P;\xb,{\mbox{\boldmath $p$}}_\perp)
	& = &
	\left. 	\frac{1}{2(2\pi)^3 }
	\int dx^-\,d^2{\mbox{\boldmath $x$}}_\perp
	e^{i(q^+x^- +\, {\mbox{\boldmath $p$}}_\perp \cdot {\mbox{\boldmath
	$x$}}_\perp)} \
	\bra P \vert \psib(x)  \psi(0) \vert P \ket
	\right|_{x^+ = 0} \nonumber \\
	& = &
	\left. \frac{1}{2(2\pi)^3 } \int \frac{dp^-}{2\pi}\
	\mbox{Tr}\, \left[{\cal A}_1 \slash p +
	 {\cal A}_2 \slash P+{\cal B} \right]
	\right|_{p^+ = -q^+} \nonumber \\
	& = &
	-\frac{M}{(2\pi)^4 q^+} \int dp^2 \,\frac{{\cal B}(P,p)}{M}
	\ \equiv \
	-\frac{M\xb}{(q^+)}\,e(\xb,{\mbox{\boldmath $p$}}_\perp ),
\end{eqnarray}
thus leading to the result
\begin{equation}
	f^{[1]}(P;\xb,{\mbox{\boldmath $p$}}_\perp) =
	\frac{q^-}{Q}\,\frac{2M\xb}{Q}\,e(\xb,{\mbox{\boldmath $p$}}_\perp).
	\qquad \qquad \qquad
\end{equation}
Although the profile function $e$ appears to enter at twist 3, it will only
give contributions at ${\cal O}(Q^{-2})$ to the hadronic tensor. This is
because
$e$ is always connected with the analogous $D^{[1]}$ function, which is also
twist 3.

In the case of a quark target we can explicitly calculate ${\cal A}_i$ and $
{\cal B}$ and find
\be
	f^+(\xb, \bp_\perp )= f^-(\xb, \bp_\perp )=
	f^\perp (\xb, \bp_\perp) = e(x,\bp_\perp)=
	\delta (1-\xb)\:\delta^{2}(\bp_\perp ).
\ee

The analysis of the  $D$-profile functions can be performed along similar
lines.
First again consider $\Gamma_A$ = $\gamma^\mu$. In this case we want to
consider
 the  profile functions in the
rest system of the produced hadron $h$. For that purpose we perform the
following Lorentz transformation,
\begin{eqnarray}
	\bigl[ p_h^-,\frac{m_{h\perp}^2}{2p_h^-},
	{\mbox{\boldmath $p$}}_{h\perp} \bigr]
	& \longrightarrow &
	\bigl[ p_h^-,\frac{m_{h}^2}{2p_h^-},{\bf 0}_{\perp} \bigr], \\
	\bigl[ k^-,k^+,{\mbox{\boldmath $k$}}_{\perp} \bigr]
	& \longrightarrow &
	\bigl[ k^-,k^+ - \frac{{\mbox{\boldmath $p$}}_{h\perp} \cdot
 	{\mbox{\boldmath $k$}}_\perp}{p_h^-}
	+ \frac{{\mbox{\boldmath $p$}}_{h\perp}^2\,k^-}{2(p_h^-)^2},
	{\mbox{\boldmath $k$}}_{\perp} - \frac{k^-\,{\mbox{\boldmath
	$p$}}_{h\perp}}{p_h^-} \bigr], \\
	\bigl[ D^-,D^+,{\mbox{\boldmath $D$}}_{\perp} \bigr]
	& \longrightarrow &
	\bigl[ D^{-\prime},D^{+\prime},{\mbox{\boldmath $D$}}_{\perp}^\prime
	\bigr] \nonumber \\
	& &
	= \bigl[ D^-,D^+ - \frac{{\mbox{\boldmath $ p$}}_{h\perp} \cdot
	{\mbox{\boldmath $D$}}_\perp}{p_h^-} + \frac{{\mbox{\boldmath
	 $ p$}}_{h\perp}^2}{2(p_h^-)^2}\,D^-, {\mbox{\boldmath $D$}}_{\perp}
	- \frac{{\mbox{\boldmath $ p$}}_{h\perp}}{p_h^-}\,D^- \bigr],
\end{eqnarray}
where $m_{h\perp}^2$ = $m_h^2 + {\mbox{\boldmath $p$}}_{h\perp}^2$. The
profile functions in the primed system ($\pht'={\bf 0_\perp}$) are then
given by (note that the hadron operators now have as argument $p_h'$)
\begin{eqnarray}
	4zD^{[\gamma^-]\prime}(p_h^\prime ;z,{\mbox{\boldmath
	$k$}}_\perp^\prime)
	& = &
	 \left.
	\frac{1}{(2\pi)^3}  \int dx^+\,d^2{\mbox{\boldmath $x$}}_\perp
	e^{i(q^-x^+ -\, {\mbox{\boldmath $k$}}_{\perp}^\prime \cdot
	{\mbox{\boldmath $x$}}_\perp)} \
	\mbox{Tr}\,\bra 0 \vert \psi(x) \,a_{h}^\dagger a_{h}\,\psib(0)
	\gamma^- \vert 0 \ket  \right|_{x^- = 0 } \nonumber \\
	& = &
	 \left. \frac{1}{(2\pi)^4} \int dk^+\
	\mbox{Tr}\, \left[ \gamma^- ({\cal C}_1 \slash k' +
	{\cal C}_2 \slash p_h' +{\cal D}) \right]
	\right|_{k^- = q^-} \nonumber \\
	& = &
	 \left. \frac{2}{(2\pi)^4} \int dk^2\ \left\lgroup {\cal C}_1(p_h',k')
	+z{\cal C}_2(p_h',k') \right\rgroup
	\right|_{k^- = q^-}  \equiv \ 4z D^-(z,-z{\mbox{\boldmath
	$k$}}_{\perp}^\prime ), \\
	4zD^{[\gamma^+]\prime}(p_h';z,{\mbox{\boldmath $k$}}_\perp^\prime)
	 & = &
	\left.
	\frac{1}{(2\pi)^3}  \int dx^+\,d^2{\mbox{\boldmath $x$}}_\perp
	e^{i(q^-x^+ -\, {\mbox{\boldmath $k$}}_\perp^\prime \cdot
	 {\mbox{\boldmath $x$}}_\perp)} \
	\mbox{Tr}\,\bra 0 \vert \psi(x) \,a_{h}^\dagger a_{h}\,\psib(0)
	\gamma^+ \vert
	0 \ket  \right|_{x^- = 0} \nonumber \\
	& = &
	\left. \frac{1}{(2\pi)^4} \int dk^+\
	\mbox{Tr}\, \left[ \gamma^+ ({\cal C}_1 \slash k' +
	{\cal C}_2  \slash p_h' +{\cal D}) \right]
	\right|_{k^- = q^-} \nonumber \\
	& = &
	\left.\frac{m_h^2}{(2\pi)^4(q^-)^2} \int
	dk^2\, \left\lgroup
	\frac{(k^2+{\mbox{\boldmath $ k$}}_\perp^{\prime 2})}{m_h^2}\,
	{\cal 	C}_1(p_h^\prime,k')
	+{1 \over z}{\cal C}_2(p_h^\prime,k') \right\rgroup
	\right|_{k^- = q^-}   \nn \\
	&\equiv &
	\frac{2m_h^2}{z(q^-)^2}\, D^+ (z,-z{\mbox{\boldmath
	$k$}}_\perp^\prime ), \\
	4zD^{[\gamma_\perp]\prime}(p_h^\prime ;z,{\mbox{\boldmath
	$k$}}_\perp^\prime)
	 & = &
	\left.
	\frac{1}{(2\pi)^3}  \int dx^+\,d^2{\mbox{\boldmath $x$}}_\perp
	e^{i(q^-x^+ -\, {\mbox{\boldmath $k$}}_\perp^\prime \cdot
	 {\mbox{\boldmath $x$}}_\perp)} \
	\mbox{Tr}\,\bra 0 \vert \psi(x) \,a_{h}^\dagger a_{h}\,\psib(0)
	 \mbox{\boldmath $\gamma$}_\perp \vert
	0 \ket  \right|_{x^- = 0} \nonumber \\
	& = &
	\frac{1}{(2\pi)^4}\int dk^+\
	\mbox{Tr}\, \left. \left[\mbox{\boldmath $\gamma$}_\perp ({\cal C}_1
	\slash k' +{\cal C}_2
	\slash p_h'+ {\cal D}) \right]
	\right|_{k^- = q^-} \nonumber \\
	&=&
	\frac{2\bk'_\perp}{(2\pi)^4 \qm} \int dk^2\, {\cal C}_1(p_h^\prime,k')
	=4 \frac{\bk'_\perp}{ \qm} D^\perp (z, -z\bk'_\perp).
\end{eqnarray}
{}From this we can immediately find the values in the original frame,
\begin{eqnarray}
	D^{[\gamma^-]}(p_h;z,{\mbox{\boldmath $k$}}_\perp)
	& = &
	D^-(z, {\mbox{\boldmath $p$}}_{h\perp}- z{\mbox{\boldmath $k$}}_\perp ),
 	\\
	D^{[\gamma^+]}(p_h;z,{\mbox{\boldmath $k$}}_\perp)
	& = &
	\frac{m_h^2}{2z(q^-)^2}\,  D^+ (z, {\mbox{\boldmath $p$}}_{h\perp}-
	z{\mbox{\boldmath $k$}}_\perp ) +
	\frac{{\mbox{\boldmath $p$}}_{h\perp}^2}{2z^2(q^-)^2}\,D^-(z,
	{\mbox{\boldmath $p$}}_{h\perp}- z{\mbox{\boldmath $k$}}_\perp ) \nn \\
	&&
	\mbox{}+\frac{\pht\cdot(\bk_\perp-\pht/z)}{z^2(\qm)^2}\,D^\perp(z,
	{\mbox{\boldmath
	$p$}}_{h\perp}- z{\mbox{\boldmath $k$}}_\perp )
	\\
	D^{[\gamma_\perp]}(p_h;z,{\mbox{\boldmath $k$}}_\perp)
	& = &
	\frac{{\mbox{\boldmath $k$}}_{\perp}-\pht/z}{zq^-}\,D^\perp
	(z, {\mbox{\boldmath $p$}}_{h\perp}- z{\mbox{\boldmath $k$}}_\perp )
	+{\pht \over z \qm} D^-(z, {\mbox{\boldmath $p$}}_{h\perp}-
	z{\mbox{\boldmath $k$}}_\perp ).
\end{eqnarray}
   In
terms of $\hat T$, $\hat q$ and  the transverse momenta one has
\be
	D^{[\gamma^\mu]}(p_h;z,{\mbox{\boldmath $k$}}_\perp)
	 =   -\frac{q^+}{Q}
	\Biggl[ \left(\hat T^\mu+ \hat q^\mu\right)\,D^-
  	 + \frac{2 (k_{\perp}^\mu-p_{h\perp}^\mu/z)}{zQ}\,D^\perp
	+\frac{2p_{h\perp}^\mu}{zQ}\, D^-  +  {\cal O}\left(
	{1 \over Q^2} \right)\Biggr].
\ee

The scalar function $D^{[1]}$ introduces the additional  profile function
$E(z,{\mbox{\boldmath $p$}}_{h\perp} - z{\mbox{\boldmath $k$}}_\perp )$,
\begin{eqnarray}
	4zD^{[1]}(p_h;z,{\mbox{\boldmath $k$}}_\perp) & = & \left.
	\frac{1}{(2\pi)^3}
	\int dx^+\,d^2{\mbox{\boldmath $x$}}_\perp
	e^{i(q^-x^+ +\, {\mbox{\boldmath $k$}}_\perp \cdot {\mbox{\boldmath
	$x$}}_\perp)} \
	\mbox{Tr}\,\bra 0 \vert \psi(x) \,a_h^\dagger a_h\,\psib(0) \vert 0
	\ket  \right|_{x^- = 0} \nonumber \\
	& = &
	\left. \frac{1}{(2\pi)^4} \int dk^+\
	\mbox{Tr}\, \left[ {\cal C}_1 \slash k + {\cal C}_2 \slash p_h +
	{\cal D} \right]
	\right|_{k^- = q^-} \nonumber \\
	& = &
	\frac{2m_h}{(2\pi)^4q^-} \int dk^2\, \frac{{\cal D}(p_h,k)}{m_h}
	\ \equiv \ \frac{4m_h}{ q^-}\,E(z,  {\mbox{\boldmath
	$p$}}_{h\perp}-z{\mbox{\boldmath $k$}}_\perp ),
\end{eqnarray}
leading to the result
\begin{equation}
	D^{[1]}(p_h;z,{\mbox{\boldmath $k$}}_\perp) =
	-\frac{q^+}{Q}\,\frac{2m_h}{Q}\,E(z,{\mbox{\boldmath $p$}}_{h\perp} -
	z{\mbox{\boldmath $k$}}_\perp ).
 	\qquad \qquad \qquad
\end{equation}
We find for the quark$\rightarrow$quark  profile functions
\be
	D^-(z,\bk_\perp)= D^+(z,\bk_\perp)=D^\perp(z,\bk_\perp)=E(z,\bk_\perp)=
	\delta (1-z)\:\delta^2 (\bk_\perp).
\ee

\newpage
}
\begin{figure}
\vspace{1cm}
\caption{\sf The general hadronic tensor (a) and the hadronic tensor with quark
currents (b).}
\label{bigblobs}
\end{figure}
\begin{figure}
\vspace{1cm}
\caption{\sf The Born term in the expansion of the hadronic tensor.}
\label{born}
\end{figure}
\begin{figure}
\vspace{1cm}
\caption{\sf One gluon contributions in the expansion of the hadronic tensor.
Only quark contributions are shown.}
\label{gauge}
\end{figure}
\begin{figure}
\vspace{1cm}
\caption{\sf{ Definition of scattering plane and momenta in $lH \rightarrow
l'hX$.}}
\label{angles}
\end{figure}
\begin{figure}
\vspace{1cm}
\caption{\sf Gluon contributions giving rise to logarithms.}
\label{logs}
\end{figure}
\begin{figure}
\vspace{1cm}
\caption{\sf Rules for the calculation of the diagrams.}
\label{rules}
\end{figure}
\begin{figure}
\vspace{1cm}
\caption{\sf A general multi-parton contribution.}
\label{factorization}
\end{figure}

\end{document}